\documentclass[aps,prx,twocolumn,superscriptaddress,showpacs]{revtex4-1}
\usepackage{bm,graphicx,amsmath,amssymb,color,placeins,tikz,pgflibraryarrows,braket,colortbl,xcolor,mathtools}
\def\k{\kappa}
\def\a{\alpha}

\def\DD{\Delta}

\def\btau{{\bm \tau}}

\def\bG{{\bf G}}
\def\bq{{\bf q} }
\def\bQ{{\bf Q} }
\def\bR{{\bf R} }
\def\bt{{\bm \tau}}
\def\bDt{{\DD \boldsymbol \tau} }

\def\<{\langle}
\def\>{\rangle}

\let\hide\iffalse
\usepackage[caption=false]{subfig}

\begin{document}

\title{Multi-phonon diffuse scattering in solids from first-principles: \\ 
Application to layered crystals and 2D materials}

\author{Marios Zacharias}
\email{marios.zacharias@cut.ac.cy}
\affiliation{Department of Mechanical and Materials Science
  Engineering, Cyprus University of Technology, P.O. Box 50329, 3603
  Limassol, Cyprus}
\author{H\'el\`ene Seiler}
\affiliation{Fritz-Haber-Institut, Physical Chemistry Department, Berlin, 14195, Germany}
\author{Fabio Caruso}
\affiliation{Institut f\"ur Theoretische Physik und Astrophysik, Christian-Albrechts-Universit\"at zu Kiel, D-24098 Kiel, Germany}
\author{Daniela Zahn}
\affiliation{Fritz-Haber-Institut, Physical Chemistry Department, Berlin, 14195, Germany}
\author{Feliciano Giustino}
\affiliation{ Oden Institute for Computational Engineering and Sciences, The University of Texas at Austin,
Austin, Texas 78712, USA
}%
\affiliation{Department of Physics, The University of Texas at Austin, Austin, Texas 78712, USA}
\author{Pantelis C. Kelires}
\affiliation{Department of Mechanical and Materials Science
  Engineering, Cyprus University of Technology, P.O. Box 50329, 3603
  Limassol, Cyprus}
\author{Ralph Ernstorfer}
\email{ernstorfer@fhi-berlin.mpg.de}
\affiliation{Fritz-Haber-Institut, Physical Chemistry Department, Berlin, 14195, Germany}

\date{\today}

\begin{abstract}

Time-resolved diffuse scattering experiments have gained increasing attention  
due to their potential to reveal non-equilibrium dynamics of crystal lattice 
vibrations with full momentum resolution. Although progress has been made in 
interpreting experimental data on the basis of one-phonon scattering, understanding 
the role of individual phonons can be sometimes hindered by multi-phonon excitations. 
In Ref.~[{\it arXiv:2103.10108}] we have introduced a rigorous 
approach for the calculation of the all-phonon inelastic scattering intensity of 
solids from first-principles. In the present work, we describe our implementation 
in detail and show that multi-phonon interactions are captured efficiently by 
exploiting translational and time-reversal symmetries of the crystal. We demonstrate 
its predictive power by calculating the scattering patterns of monolayer molybdenum 
disulfide (MoS$_2$), bulk MoS$_2$, and black phosphorus (bP), and we obtain excellent 
agreement with our measurements of thermal electron diffuse scattering. 
Remarkably, our results show that multi-phonon excitations dominate in bP across multiple 
Brillouin zones, while in MoS$_2$ they play a less pronounced role. We expand our analysis for 
each system and examine the effect of individual atomic and interatomic vibrational motion 
on the diffuse scattering signals. We further demonstrate the high-throughput capability 
of our approach by reporting all-phonon scattering maps of 2D MoSe$_2$, WSe$_2$, WS$_2$, 
graphene, and CdI$_2$, rationalizing in each case the effect of multi-phonon processes. 
As a side point, we show that the special displacement method reproduces the thermally 
distorted configuration that generates precisely the all-phonon diffuse pattern. 
The present methodology opens the way for systematic calculations of the scattering 
intensity in crystals and the accurate interpretation of static and time-resolved 
inelastic scattering experiments.

\end{abstract}

\maketitle

\section{Introduction}
Nonequilibrium phenomena as diverse as phase transitions, polaron formation, electrical and thermal management in 
semiconductor devices, all derive from microscopic interactions between electrons and phonons, spins and phonons, 
as well as phonons with phonons~\cite{Waldecker_2016,Nicholson2018,Na2019,Caruso_2021}. 
Our understanding of such phenomena hinges on the development of joint experimental and theoretical tools which 
can access these interactions at the mode-resolved level with sufficient temporal resolution. 
Towards this goal, exciting methodological developments were recently achieved on the experimental side 
with structural probes, either using Femtosecond X-ray Diffuse Scattering or Femtosecond Electron Diffuse 
Scattering (FEDS)~\cite{Trigo2010, Trigo2013, Waldecker2017, Wall2018, Stern2018, Konstantinova_2018, 
Teitelbaum_2018, Cotret_2019, Seiler2021, Otto2021}. 
For the first time, these methods yield access to nonequilibrium phonon populations in momentum space, 
beyond the zone-center modes traditionally accessible with optical spectroscopies.

\begin{figure}[t!]
\includegraphics[width=0.46\textwidth]{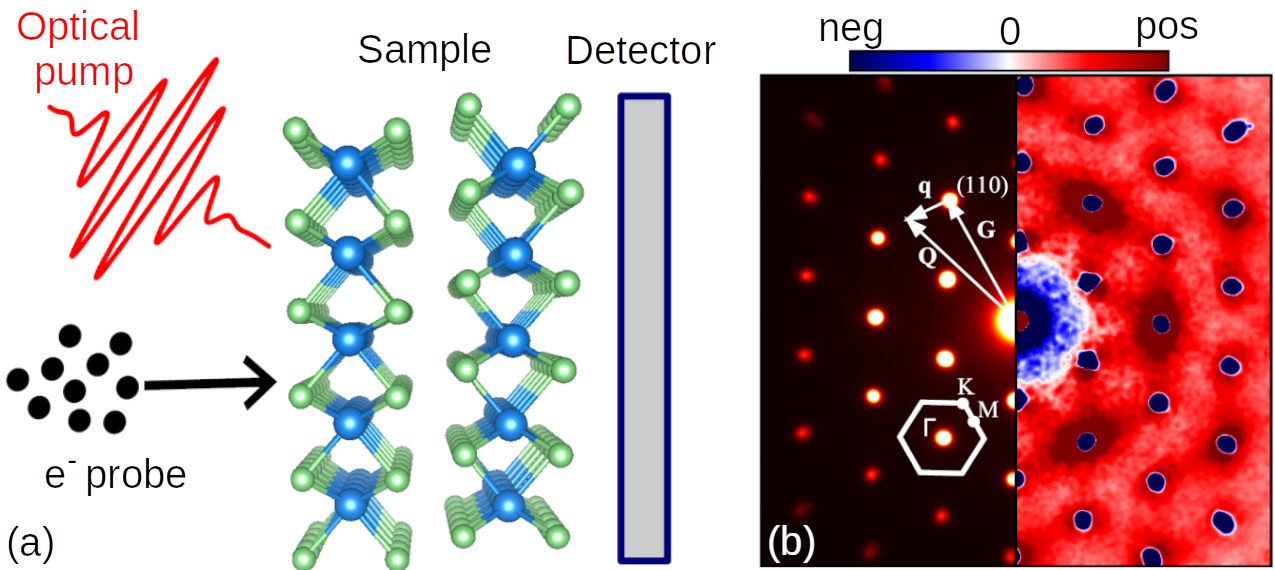}
\caption{
  (a) Schematic illustration of FEDS experiment on bulk MoS$_2$. 
   More details about the setup can be found in Sec.~\ref{sec.Experiment_Methods} and Ref.~\cite{2015Wald}.
  (b) Typical scattering pattern of bulk MoS$_2$. Left subplot shows the raw pattern as collected on the detector. 
  $\bG$ indicates the Bragg peak vector (110), $\bq$ the reduced phonon wavevector and $\bQ = \bG + \bq$ the scattering vector. 
  A hexagonal Brillouin zone with the high-symmetry points $\Gamma$, K, and M are also shown. 
  Right subplot shows the difference scattering pattern 
  $\Delta I(\bQ, 100 {\rm \, ps}) = I(\bQ, 100 {\rm \, ps} ) - I(\bQ, t < t_0)$, 
  where $I(\bQ, t < t_0)$ is the average intensity prior to photoexcitation. 
  \label{fig1}}
\end{figure}

In these experiments, the observable depends on the temporal evolution of the 
scattering intensity $I(\bQ, t)$, where $\bQ$ is an arbitrary scattering wavevector 
determined by the difference in momentum of the incident and scattered radiation. 
The key information obtained %in a time-resolved diffuse scattering measurement 
is the changes in the diffracted intensities, as they reflect how different phonons 
get populated as a function of time $t$. In FEDS, these changes are visualized by computing 
the difference scattering pattern $\Delta I(\bQ, t)$~\cite{Zahn_2020,Seiler2021}.  
In Figs.~\ref{fig1}(a) and~(b) we present a schematic illustration of FEDS and an typical $\Delta I(\bQ, t)$ % difference diffraction pattern
of bulk molybdenum disulfide (MoS$_2$). The left subplot in  Fig.~\ref{fig1}(b) simply shows the intensity 
as collected on the detector. Each $\bQ$ on this pattern can be expressed as a summation of a
Bragg peak vector $\bG$ (centers of the Brillouin zones), and reduced phonon wavevectors $\bq$. 
The right subplot shows $\Delta I(\bQ, 100 {\rm \, ps})$ and displays a hot, but quasi-thermalized
distribution of phonons in the MoS$_2$ sample. The blue/red features represent a decrease/increase in the signal 
due to Bragg/diffuse scattering. 
The larger the intensity of the red features indicates regions of the reciprocal space with higher 
phonon scattering probability. Recent works have shown that $\Delta I(\bQ, t)$ can change profoundly 
and qualitatively as time evolves, reflecting non-thermal lattice dynamics~\cite{Stern2018, Waldecker2017, Seiler2021}. 
Phonon populations typically evolve towards a hot, but thermal distribution [e.g. right subplot of Fig.~\ref{fig1}(b)]
with a highly material-specific timescale.

Although FEDS measurements possess a wealth of information, data interpretation is rather complex due to the 
energy-integrated nature of the experiment and the multiple scattering phenomena involved. 
Therefore, before analysing the highly non-equilibrium phonon distributions, it is necessary 
to fully understand thermal diffuse scattering, i.e., inelastic scattering induced by phonons, 
using first-principles calculations. %One possible analysis route, 
Recent first-principles calculations of  phonon-diffuse scattering~\cite{Konstantinova_2018,
Krishnamoorthy_2019,Cotret_2019,Bernardi_2021,Maldonado_2020,Seiler2021, Otto2021} 
rely on the quantum theory of the one-phonon structure factor~\cite{Grosso_Pastori_book,Zhong_Lin_Wang}. 
%%%%%%%%%%%%%%%%%%%
Despite their great success in explaining some of the main features 
in the diffuse pattern, one-phonon interactions are considered inadequate
to explain scattering signals at large $|\bQ|$ and/or high temperatures~\cite{Zhong_Lin_Wang,arXiv:2103.10108}.
In these cases, the intensity contributed by multi-phonon scattering can become comparable with, 
and even larger than, that of one-phonon excitations.
A multi-phonon process occurs when the momentum transfer to the beam in a single scattering event %excites more than one phonons, 
%and the momentum transfer to the crystal 
is specified by more than one phonons. % involved in the process.
This principle is well described in the 
literature~\cite{Sjolander_1958,Dawidowski_1998,Baron_2007,Kuroiwa_2008,Baron2014,Wehinger_2017}.
Other mechanisms that contribute to diffuse signals
are multiple interactions (i.e. more than one electron scattering 
events~\cite{Zhong_Lin_Wang}), inelastic scattering on plasmons and defects,  
or surface imperfections making the role of multi-phonon scattering inconclusive~\cite{Vallejo_2018}.
This situation highlights the need for computational tools that directly probe multi-phonon contributions
and, in essence, go a step forward to extract phonon population dynamics across the entire 
Brillouin zone~\cite{Cotret_2019}.

In the parallel paper, Ref.~\cite{arXiv:2103.10108}, we have introduced a methodology for the calculation of 
the all-phonon scattering in solids, 
which enables us to single out the contribution of phonon interactions, 
and thus isolate their scattering signatures. In the present work, we further validate our implementation by first calculating 
one- and multi-phonon scattering patterns of monolayer MoS$_2$. Using the same system, we also 
demonstrate that the special displacement method (SDM)~\cite{Zacharias_2016,Zacharias_2020} can provide an alternative route
for the assessment of all-phonon contributions. We then apply our technique for the calculation of bulk MoS$_2$ 
and black phosphorus (bP) scattering patterns and obtain excellent agreement with experiment. 
Importantly, our results reveal that multi-phonon interactions are more manifested in bP than in MoS$_2$. 
We also demonstrate the efficiency of our technique by evaluating 
phonon-induced scattering patterns of several 2D materials.
Although this work focuses on a comparison between theory and FEDS measurements, we emphasize that the developments 
presented here  are fully applicable to X-ray, or neutron, diffuse scattering.

The organization of the manuscript is as follows: in Secs.~\ref{sec.Theory_1} and~\ref{sec.Theory_2} 
we describe the theory of quantum mechanical scattering in solids and derive the main equations 
used to evaluate the respective phonon contributions. In Sec.~\ref{ZG_theory} and Appendix~\ref{app.equiv}
we demonstrate that SDM can serve as an equivalent, but different, approach for calculating the all-phonon scattering intensity. 
In Sec.~\ref{Einstein_Model} we describe the Einstein model for diffuse scattering. 
Sections~\ref{sec.Experiment_Methods} and~\ref{sec.Theory_Methods} report all experimental and 
computational details of the measurements and calculations performed in this work. 
In Sec.~\ref{sec.results} we present our results for several 2D materials, bulk 
MoS$_2$, and bP. Specifically, in Sec.~\ref{sec.res_2D_MoS2}
we report scattering intensity calculations of 2D MoS$_2$ using the exact theory, special displacements, 
and the Einstein model. In Secs.~\ref{sec.res_bulk_MoS2} and~\ref{sec.res_bulk_bP} we report the phonon 
scattering intensities of bulk MoS$_2$ and bP, respectively, and compare our calculations of the difference 
patterns with experiment. The results are accompanied by an analysis of the multi-phonon contribution across 
multiple Brillouin zones, as well as of the scattering signatures of individual atomic and interatomic
thermal motion. In Sec.~\ref{sec.2D_materials} we further validate 
our approach on monolayers MoSe$_2$, WSe$_2$, WS$_2$, graphene, and CdI$_2$. 
Our conclusions and outlook are presented in Sec.~\ref{sec.Conclusions}. 

\section{Theory} \label{sec.Theory} 

In this section we present the theoretical framework underpinning the evaluation of multi-phonon scattering intensity.
Starting from the Laval-Born-James~\cite{Laval_1939,Born_1942,James_1948} theory, we derive the 
zero-phonon, one-phonon, and all-phonon scattering intensities in the harmonic approximation. 
We also demonstrate that exact phonon-diffuse patterns can be evaluated using SDM
in the limit of dense Brillouin sampling. 

We stress that all subsequent expressions apply for electron, X-ray, and neutron diffuse 
scattering under the assumption of the kinematic limit~\cite{Zhong_Lin_Wang}. 
In this limit, also known as the Born-approximation, the Lippmann-Schwinger quantum formulation 
for particle scattering~\cite{Lippman_1950} is truncated up to the first order in the interaction 
potential, thus neglecting multiple scattering events. That amounts to assume weak 
interactions where the incident beam is scattered only once by the crystal.

In the following we adopt a similar notation as in Ref.~[\onlinecite{Giustino_2017}].

\subsection{Scattering intensity}  \label{sec.Theory_1} 
In the adiabatic formulation and kinematic limit of the quantum mechanical 
scattering theory, originally developed by Laval~\cite{Laval_1939}, Born~\cite{Born_1942}, and James~\cite{James_1948} (LBJ), 
the intensity of the wave scattered by the atoms in a crystal is given by~\cite{Maradudin_Weiss}:
\begin{eqnarray}\label{eq.intensity}
   I_{\alpha n , \beta m}(\bQ) = \bigg| \bra{X_{\alpha n}} \sum_{p \k} f_\k (\bQ) 
   e^{i \bQ \cdot [\bR_p + \bt_\k + \bDt_{p\k }]} \ket{X_{\beta m}}   \bigg|^2. \nonumber
\\ 
\end{eqnarray}
Here the many-body electron-nuclear system is described in terms of the Born-Huang expansion~\cite{born1954},
with $\ket{X_{\alpha n}}$ and $\ket{X_{\beta m}}$ representing the initial $n$ and final $m$ Born-Oppenheimer vibrational states 
which are associated with electronic states denoted by the Greek indices $\alpha$ and $\beta$.
%of the crystal lattice undergoing a scattering event, with the Greek indices representing the associated electronic states. 
The summations run over all atoms $\k$ in the unit cell and over all $p$ indices of the direct lattice vectors
$\bR_p$. The lattice vectors define a Born-von K\'arm\'an supercell which contains $N_p$ unit cells. 
The atomic scattering amplitude is denoted by $f_\k (\bQ)$ and is evaluated at the scattering vector $\bQ$. 
The displacement vector of the atom $\k$ from its equilibrium position vector $\bt_\k$
is represented by $\bDt_{p\k }$.
For generality and brevity reasons, the intensity $I(\bQ)$ is expressed in  
scattering units depending on the probe-sample interaction~\cite{Maradudin_Weiss}. 

If we set the initial and final electrons in their Born-Oppenheimer ground state, i.e. $\alpha = \beta = 0$, 
perform the summation over all final vibrational states of the scatterer in Eq.~\eqref{eq.intensity}, 
and use the closure relationship $\sum_m \ket{X_{0 m}} \bra{X_{0 m}} = \mathbb{I}$, 
 then we obtain: 
\begin{eqnarray}\label{eq.intensity_2}
   I_{0 n}(\bQ) =  \bra{X_{0 n}} I^{\{\tau\}}(\bQ) \ket{X_{0 n}},
\end{eqnarray}
where 
\begin{eqnarray}\label{eq.intensity_3}
 I^{\{\tau\}}(\bQ) =   \bigg| \sum_{p \k} f_\k (\bQ) 
   e^{i \bQ \cdot [\bR_p + \bt_\k + \bDt_{p\k }]}  \bigg|^2  %= \\
\end{eqnarray}
represents the scattering intensity arising from an instantaneous atomic configuration
defined by the set of atomic displacements \{$\bDt_{p \k}$\}. 
We note that setting the electronic states at their ground level is justified for a system 
at thermal equilibrium before, and after, diffraction~\cite{Born_1942}.  

The LBJ scattering intensity at finite temperature $T$ is obtained from Eq.~\eqref{eq.intensity_2}
by taking the ensemble average over all possible configurations of the nuclei. That is:
\begin{equation}\label{eq.intensity_T}
  I(\bQ,T) = \frac{1}{Z} \sum_n \exp(-E_{0 n}/k_{\rm B}T)\, I_{0 n}(\bQ),
 \end{equation}
where $E_{0 n}$ stands for the energy of the nuclear state $ \ket{X_{0 n}}$,
  $Z = \sum_n \exp(-E_{0 n}/k_{\rm B}T)$ is the canonical partition function, and $k_{\rm B}$ is 
the Boltzmann constant. The above relation can also be recognized as the 
Williams-Lax~\cite{Williams_1951,Lax_1952} thermal average of the scattering intensity. 
This can be understood by writing the scattering intensity as a Fermi Golden rule [similar to
Eq.~(3) of Ref.~\cite{Zacharias_2020}], consider no electronic excitations, and integrate over
the energy transfer to the crystal~\cite{Grosso_Pastori_book}. 
An alternative interpretation is that Eq.~\eqref{eq.intensity_T} represents, essentially, the 
static limit of the dynamic structure factor~\cite{Baron2014}, accounting for an average of
all initial and final vibrational states accessible at thermal equilibrium.

\subsection{Exact evaluation: Zero-phonon, one-phonon and all-phonon scattering intensities} \label{sec.Theory_2}

Now, starting from Eq.~\eqref{eq.intensity_T} and employing the harmonic approximation,
we derive the formulas of the zero-phonon (elastic scattering), one-phonon and multi-phonon (inelastic scattering)
contributions. 
To this aim we adopt the normal mode coordinate formalism and first write the atomic displacement vector as:
\begin{equation}\label{eq.dtau}
    \bDt_{ p\k} = \bigg(\frac{M_0}{N_p M_\k}\bigg)^{1/2} \sum_{ \bq \nu} e^{i\bq \cdot {\bf R}_p} 
      {\bf e}_{\k,\nu} (\bq ) z_{\bq  \nu}, 
\end{equation}
where $z_{\bq  \nu}$ are the complex-valued normal coordinates associated with the mode
of reduced wavevector $\bq$ and branch index $\nu$, 
$M_\k$ is the mass of the $\k$th atom, and $M_0$ is the atomic mass unit. 
The phonon polarization vector of the normal mode is denoted 
as $ {\bf e}_{\k,\nu} (\bq )$  with Cartesian components $ e_{\k\a,\nu} (\bq)$. 

In the framework of the harmonic approximation, the nuclear wavefunction
$ \ket{X_{0 n}}$ is expressed as a Hartree product of uncoupled quantum harmonic oscillators  
and the nuclear energy $E_{0 n}$ as a summation over the associated energy quanta.
Writing the harmonic oscillators in terms of Hermite polynomials and 
employing Mehler's sum rule~\cite{Watson_1933}
leads to the following integral form for the LBJ scattering intensity~\cite{Patrick_2014,Zacharias_2020}: 
\begin{eqnarray}\label{eq_SMD_2}
     \hspace{-5pt} I(\bQ,T)  &=& \Braket{ I^{\{\tau\}}(\bQ) }_T \\ & =&
     \prod_{\bq\nu} \!\int\! \frac{dz_{\bq \nu} }{\pi u^2_{\bq  \nu}} 
    e^{-| z_{\bq \nu}|^2/ u^2_{\bq  \nu}} I^{\{\tau\}}(\bQ). \nonumber  %\hspace{10pt} \\
 \end{eqnarray}
Here $\Braket{.}_T$ represents the ensemble thermal average which is taken as a multidimensional Gaussian integral 
over the normal coordinates in the same way as a Williams-Lax observable in the harmonic approximation~\cite{Zacharias_2015}. 
The widths of the Gaussians are determined by the mode-resolved
 mean-square displacement of the atoms at temperature $T$: 
\begin{eqnarray}\label{eq_SDM_3}
u^2_{\bq \nu}  = \frac{\hbar}{2M_0 \omega_{\bq \nu}}[2n_{\bq  \nu}(T) + 1 ], 
\end{eqnarray}
where $n_{\bq  \nu}(T)$ represents the Bose-Einstein occupation 
of the phonon with frequency $\omega_{\bq \nu}$ at thermal equilibrium, but can depart
significantly from this value under nonquilibrium conditions~\cite{Caruso_2021,Seiler2021}. We note that Eq.~\eqref{eq_SDM_3} 
is indefinite for the zero-frequency translational modes (accoustic modes at $\Gamma$).
These modes do not impose any change on the properties of the lattice and thus 
the associated mean-square displacement can be set to zero. 

The exact expression for the calculation of the temperature-dependent scattering intensity 
is obtained with the aid of the Bloch identity~\cite{Grosso_Pastori_book}: 
 \begin{eqnarray}\label{eq_key}
   \Braket{e^{i \bQ \cdot  \bDt_{p\k } }}_T = 
            e^{-\frac{1}{2}  
  \Braket{ \big( \bQ \cdot  \bDt_{p\k } \big)^2 }_T}. 
\end{eqnarray}
Hence, combining Eqs.~\eqref{eq.intensity_3} and~\eqref{eq_SMD_2} yields:
 \begin{eqnarray}\label{eqa1.3}
        I(\bQ,T)  
           & = &  \sum_{pp'} \sum_{\k \k'} f_\k (\bQ) f^*_{\k'} (\bQ) 
            e^{i \bQ \cdot [\bR_p - \bR_{p'} + \bt_\k - \bt_{\k'} ]} \nonumber \\
            &\times&  e^{-\frac{1}{2}  
  \Braket{ \big\{ \bQ \cdot  (\bDt_{p\k } -  \bDt_{p'\k' } ) \big\}^2 }_T}. 
\end{eqnarray}
By replacing now $\bDt_{p\k }$ with the normal-coordinate transformation of Eq.~\eqref{eq.dtau},
considering translational invariance of the lattice, and using 
the identity $\Braket{z_{\bq  \nu} z^*_{\bq'  \nu'}}_T = u^2_{\bq \nu} \, \delta_{\bq \bq', \nu \nu'}$, we obtain 
the following compact form for the LBJ (or {\it all-phonon}) scattering intensity~\cite{Xu2005}: 
\begin{eqnarray}\label{eqa1.7}
I_{\rm all}(\bQ,T) &=& N_p  \sum_{p} \sum_{\k \k'} f_\k (\bQ) f^*_{\k'} (\bQ)
    e^{ i \bQ \cdot [\bR_p + \bt_\k - \bt_{\k'} ] } \nonumber 
    \\ &\times&  e^{-W_{\k} (\bQ,T)} \, e^{-W_{\k'} (\bQ,T)} \, e^{ P_{p,\k\k'} (\bQ,T)}.
\end{eqnarray}
We emphasize that this formula is identical to the Van Hove's dynamical 
structure factor for inelastic scattering~\cite{Van_Hove_1954} when integrated over 
phonon energies, thus, accounting precisely for all phonon absorption and emission processes. 
Here, the exponent of the Debye-Waller factor is defined as:
\begin{eqnarray}\label{eqa1.8}
 -W_{\k} (\bQ,T) &=& -\frac{M_0}{N_p M_\k }  \sum_{\bq \in \mathcal{B},\nu }  
\big|\bQ \cdot {\bf e}_{\k,\nu} (\bq )\big|^2  u^2_{\bq \nu} \\ 
&-& \frac{M_0}{2 N_p M_\k }  \sum_{\bq \in \mathcal{A},\nu }  
\big|\bQ \cdot {\bf e}_{\k,\nu} (\bq )\big|^2  u^2_{\bq \nu}, \nonumber
\end{eqnarray}
and the exponent of the phononic factor as:
\begin{widetext}
\begin{eqnarray}\label{eqa1.8_b}
 P_{p,\k\k'} (\bQ,T)  &=& \frac{2 M_0 N^{-1}_p}{\sqrt{M_\k M_{\k'}}} \sum_{\bq \in \mathcal{B},  \nu }   u^2_{\bq \nu} 
\text{Re}\bigg[ \bQ \cdot {\bf e}_{\k,\nu} (\bq) \bQ \cdot {\bf e}^{*}_{\k',\nu} (\bq) e^{i\bq \cdot {\bf R}_p} \bigg] \nonumber \\
  &+& \frac{ M_0 N^{-1}_p}{\sqrt{M_\k M_{\k'}}} \sum_{\bq \in \mathcal{A},  \nu }   u^2_{\bq \nu}
 \bQ \cdot {\bf e}_{\k,\nu}(\bq) \bQ \cdot {\bf e}_{\k',\nu} (\bq) \cos(\bq \cdot {\bf R}_p).
\end{eqnarray}
\end{widetext}
The summations are restricted to: (i) the group $\mathcal{B}$ 
containing phonons with wavevectors that lie in the Brillouin zone and
are not time-reversal partners, and (ii) the group $\mathcal{A}$ containing phonons that remain invariant 
under time-reversal~\cite{Zacharias_2020}. 
Re[.] represents the function that returns the real part of the argument 
inside the square brackets.
Combining the partitioning of phonons in groups $\mathcal{A}$ and $\mathcal{B}$ with the use of translational symmetry
of the crystal enables the efficient calculation of the all-phonon diffuse scattering intensity. 
This aspect is central in this manuscript and allows for the rapid assessment of multi-phonon excitations.
The summations over different pairs of atoms in Eq.~\eqref{eqa1.7} can be conveniently partitioned 
into different parts to examine individual ($\k = \k'$) and distinct ($\k \neq \k'$) scattering
contributions~\cite{Van_Hove_1954}. 

Physically, the Debye-Waller factor, $e^{-W_{\k}}$, determines the attenuation of the scattering intensity 
at temperature $T$ owing to the vibrational motion of atom $\k$.
The phononic factor, $e^{P_{p,\k\k'}}$, includes all-phonon
contributions to diffuse scattering associated with the individual or 
combined thermal motion of atoms $\k$ and $\k'$ in unit cell $p$.
For example, the zero-phonon, $I_0$, and one-phonon, $I_1$, contributions  
are obtained by retaining the zeroth and first-order terms in the Taylor expansion 
of $e^{P_{p,\k\k'}}$~\cite{Grosso_Pastori_book}.
Hence, if we use the standard sum rule
 $\sum_{p} \text{exp}(i\bQ \cdot {\bf R}_p) = N_p \, \delta_{\bQ,\bG}$, where $\bG$ 
is a reciprocal lattice vector
and observe that $I_0(\bG,T) = I_0(-\bG,T) $, we can write the zero-phonon, or Bragg scattering, term as:
\begin{eqnarray}\label{eqa1.12_b}
I_0(\bQ,T) &=&   N_p^2 \sum_{\k \k'} f_\k (\bQ) f^*_{\k'} (\bQ)
    \cos\big[ \bQ \cdot (\bt_\k - \bt_{\k'})\big] \nonumber  \\ 
&\times& e^{-W_{\k} (\bQ,T)} e^{-W_{\k'} (\bQ,T)}  \delta_{\bQ,\bG}. 
\end{eqnarray}
This expression is directly related to Laue's interference condition
and has very sharp maxima whenever $\bQ = \bG$, and reduces to zero otherwise.

Similarly to the zero-phonon term, one can obtain a compact formula for the one-phonon contribution 
to the scattering intensity by following a straightforward, but more lengthy, derivation. 
The final result is: %for $\bQ \equiv \bG + \bq \in B $ is: 
\begin{eqnarray}\label{eqa1.13}
&\,& I_1(\bQ,T) =  M_0 N_p \sum_{\k \k'} f_\k (\bQ) f^*_{\k'} (\bQ)
\frac{ e^{-W_{\k} (\bQ,T)}
   e^{-W_{\k'} (\bQ,T)}}{\sqrt{M_\k M_{\k'}}} \nonumber  \\ &&
 \times   \sum_{\nu } \, \text{Re} \Big[ \bQ \cdot {\bf e}_{\k,\nu}(\bQ) 
   \bQ \cdot {\bf e}^{*}_{\k', \nu} (\bQ) e^{i\bQ \cdot [\bt_{\k'} - \bt_{\k}]} \Big] u^2_{\bQ  \nu}. 
\end{eqnarray}
One can continue the analysis and derive explicit expressions for the intensity 
of each higher-order process. Notably, each expression is positive definite and, thus, 
multi-phonon scattering contributes constructively so that $I_{\rm all }(\bQ,T) \geq I_0(\bQ,T) + I_1(\bQ,T) $. 

\subsection{All-phonon scattering intensity using the special displacement method} \label{ZG_theory}

Recently, it has been shown that any Williams-Lax thermal average in the form of Eq.~\eqref{eq_SMD_2} 
can be evaluated using the special displacement method (SDM)
developed by Zacharias and Giustino (ZG)~\cite{Zacharias_2016,Zacharias_2020}.  
SDM amounts to applying ZG displacements on the nuclei away from
their equilibrium positions given by~\cite{Zacharias_2020}:
 \begin{eqnarray}\label{eq.realdtau_method00}
   \DD\btau^{\rm ZG}_{ p\k} &=&  \sqrt{\frac{M_0}{N_p M_\k}}\Bigg[ \!\sum_{\bq \in \mathcal{B}, \nu}\! 
   S_{\bq  \nu} u_{\bq \nu} 2 \,{\rm Re}
   \Big[ e^{i\bq \cdot {\bf R}_p} {\bf e}_{\k,\nu} (\bq ) \Big] \nonumber \\
    &+&  \!\sum_{\bq \in \mathcal{A}, \nu}\! 
   S_{\bq  \nu} u_{\bq \nu} 
   \cos(\bq \cdot {\bf R}_p) {\bf e}_{\k,\nu} (\bq ) \Bigg] . 
 \end{eqnarray}
In the above relation the amplitudes of the normal coordinates entering Eq.~\eqref{eq.dtau} 
are set to $|z_{\bq \nu}| = u_{\bq \nu}$, and their signs to $S_{\bq  \nu}$.
For practical calculations, the choice of signs is made such that the following function is minimized: 
%\begin{widetext}
\begin{eqnarray}\label{eq.minimiz_function}
E(\{S_{\bq  \nu}\}&,&T) = \\ \sum_{\substack{\k\a \\ \k'\a' }} &\bigg|& \sum_{\substack {\bq \in B \\ \nu < \nu'}} 
 \text{Re} [  e_{\k\a,\nu}(\bq)  e^{*}_{\k'\a',\nu'}(\bq)] u_{\bq \nu} u_{\bq \nu'}  S_{\bq  \nu}  S_{\bq  \nu'} \nonumber \\ 
&+& \sum_{\substack {\bq \in A \\ \nu < \nu'}} 
  e_{\k\a,\nu}(\bq)  e^*_{\k'\a',\nu'}(\bq) u_{\bq \nu} u_{\bq \nu'}   S_{\bq  \nu}  S_{\bq  \nu'} \nonumber
\bigg|
\end{eqnarray}
%\end{widetext}
The above formula reduces exactly to zero in the limit of dense Brillouin zone sampling, since all 
summands inside the modulus remain nearly the same and have opposite sign for adjacent ${\bf q}$-points~\cite{Zacharias_2020}. 
More details about the allocation of the signs $S_{\bq  \nu}$, 
as well as the ordering of phonons for the construction of ZG displacements are given in Sec.~\ref{sec.Theory_Methods}. 
Minimization of Eq.~\eqref{eq.minimiz_function} guarantees 
that: (i) the  nonperturbative error in the calculation of the temperature-dependent 
observable is eliminated, and (ii) the quantum mechanical anisotropic displacement tensor of the atoms, 
defined as~\cite{Maradudin_Weiss_1963}
  \begin{eqnarray}\label{eq.mean-square_disp}
   U_{\k,\a\a'}(T) &=& \frac{2M_0}{N_p M_\k} \sum_{\bq \in \mathcal{B}, \nu} 
   \text{Re}[e_{\k \a, \nu } (\bq) e^{*}_{\k \a',\nu } (\bq) ]\,\, u^2_{\bq \nu} \nonumber \\
     &+& \frac{M_0}{N_p M_\k} \sum_{\bq \in \mathcal{A}, \nu} 
   e_{\k \a, \nu } (\bq) e_{\k \a',\nu } (\bq) \,\, u^2_{\bq \nu},
 \end{eqnarray}
is recovered. This quantity also determines the thermal ellipsoids of 
the crystal and its diagonal elements are closely related to the exponent of the Debye-Waller factor
given by Eq.~\eqref{eqa1.8}. 

The calculation of the scattering intensity at finite temperatures using
SDM requires to simply set $\bDt_{p\k}= \bDt^{\rm ZG}_{p\k}$ in Eq.~\eqref{eq.intensity_3},
and thus calculate Eq.~\eqref{eq.intensity_T} for a single distorted configuration. That is: 
 \begin{eqnarray}\label{eq.ZG_struct_factor}
  I_{\rm ZG}(\bQ,T) %&=& I^{\{\tau^{\rm ZG}\}}(\bQ) \nonumber \\
  &=&   \bigg| \sum_{p \k} f_\k (\bQ) 
   e^{i \bQ \cdot \big[\bR_p + \bt_\k + \bDt^{\rm ZG}_{p\k } \big]}  \bigg|^2. 
\end{eqnarray}
The proof that the Williams-Lax thermal average of a generic observable can be evaluated using the ZG displacements
is provided in Ref.~\cite{Zacharias_2020}. 
In Appendix~\ref{app.equiv}, we demonstrate, using a different approach,   
that Eq.~\eqref{eq.ZG_struct_factor} is equivalent to Eq.~\eqref{eqa1.7},  
as long as Eq.~\eqref{eq.minimiz_function} is minimized. 
This finding reinforces the concept that nuclei positions defined by ZG displacements can describe
accurately thermal disorder in solids and, here, can be viewed as {\it the collection of scatterers that best 
reproduce the diffuse scattering intensity}. 

\subsection{Scattering intensity using the Einstein model} \label{Einstein_Model}

For an Einstein solid, the scattering intensity can be evaluated by assuming that all atoms 
vibrate independently and with the same frequency~\cite{Hall_1965_b}. These approximations allow one to 
replace: (i) the mode-resolved mean-square displacement of the atoms $u^2_{\bq \nu}$ by a constant
$u^2_{\rm E} = \hbar / (2M_0 \omega_{\rm E})[2n_{{\rm E}}(T) + 1 ]$, where $\omega_{\rm E}$ 
is the average phonon frequency of the crystal and $n_{{\rm E}}$ the associated Bose-Einstein occupation,
and (ii) the phonon polarization vectors $ {\bf e}_{\k,\nu} (\bq )$ with a normalized isotropic eigenvector~\cite{Hall_1965}. 
Applying (i) and (ii) to Eq.~\eqref{eqa1.7}, the scattering intensity within the Einstein model reads:
\begin{eqnarray}\label{eqa1.7_Einstein}
I_{\rm E}(\bQ,T) &=& N^2_p  \sum_{\k \k'} f_\k (\bQ) f^*_{\k'} (\bQ)
    \cos\big[ \bQ \cdot ( \bt_\k - \bt_{\k'} ) \big] \nonumber 
    \\ &\times&  e^{-C_{\k\k}(\bQ,T) } \, e^{ -C_{\k'\k'}(\bQ,T) }  
   \delta_{\bQ,\bG}  \\
  &+& N_p  \sum_{\k \k'} f_\k (\bQ) f^*_{\k'} (\bQ)
    \cos\big[ \bQ \cdot ( \bt_\k - \bt_{\k'} ) \big] \nonumber 
    \\ &\times&  e^{-C_{\k\k}(\bQ,T) } \, e^{ -C_{\k'\k'}(\bQ,T) } 
    \,  \Big[ e^{2 C_{\k\k'}(\bQ,T)} - 1   \Big], \nonumber 
\end{eqnarray}
where $C_{\k\k'}(\bQ,T) = M_0 u^2_{E} \bQ^2 / \sqrt{4 M_\k M_{\k'}}$. 
The first and second summations represent the elastic and inelastic terms, respectively. 
The above oversimplified expression provides a quick estimate of the contribution of the first and higher 
order excitations based on the power series expansion of $e^{2 C_{\k\k'}(\bQ,T)}$. 
For example, keeping terms up to the first order in $C_{\k\k'}(\bQ,T)$ yields the Einstein model's 
analogue of Eq.~\eqref{eqa1.13}.
% of the dimensionless quantity $C_{\k\k'}(\bQ,T)$. 

\section{Methods} \label{sec.Methods}

\subsection{Experiment} \label{sec.Experiment_Methods}

The FEDS measurements are performed in transmission using the compact diffractometer described in detail elsewhere 
\cite{2015Wald}. Briefly, the output of a femtosecond laser system
(Astrella, Coherent, 4 kHz, pulse duration 50 fs) is split into a pump arm 
and a probe arm. A commercial optical parametric amplifier is used to generate pump pulses with tunable wavelength.
The electron probe is generated from two-photon absorption of around $500$~nm photons obtained 
from a home-built non-collinear optical parametric amplifier (NOPA) and subsequent 
photoemission from a gold photo-cathode. The photo-emitted electron 
bunches are accelerated towards the anode to reach $60$-$90$~keV as they exit the gun. Each electron 
bunch is estimated to contain $\simeq 10^3$ electrons.
Scattering patterns are recorded with a phosphor screen fiber-coupled to a CMOS detector 
(brand TVIPS, model TemCam-F416).

For sample preparation, bulk black phosphorus and MoS$_2$ crystals were purchased from \textit{HQ Graphene}. 
Free-standing thin films were obtained in both cases by mechanical exfoliation 
and subsequent transfer to TEM grids using the floating technique \cite{2007Dwyer}. 
Due to their air-sensitivity, the bP flakes were transferred to vacuum immediately 
after preparation to prevent degradation of the bulk film.

The bP data were acquired at a base temperature of $T = 100$~K, whereas the MoS$_2$ 
data were acquired at a base temperature of $T = 300$~K.
All data were processed using the open-source python module \textit{scikit-ued} \cite{RendeCotret2018}. 
In particular, a six-fold (two-fold) symmetrization operation was performed on the raw MoS$_2$ (bP) 
scattering patterns. The symmetrization operations were carried out for visualization purposes only. 
Prior to symmetrization, it was verified that the signals in corresponding Bragg orders 
(Friedel pairs) match in intensity within error, defined as the standard error of the mean signal 
over multiple independent acquisitions of the scattering pattern. In the symmetrized experimental 
patterns, we observe double peaks at large scattering vectors. These double peaks are artefactual 
and arise from magnetic field distortions of the electrons lens, which induce aberrations at 
large scattering vectors.

\begin{figure*}[hbt!]
\includegraphics[width=0.96\textwidth]{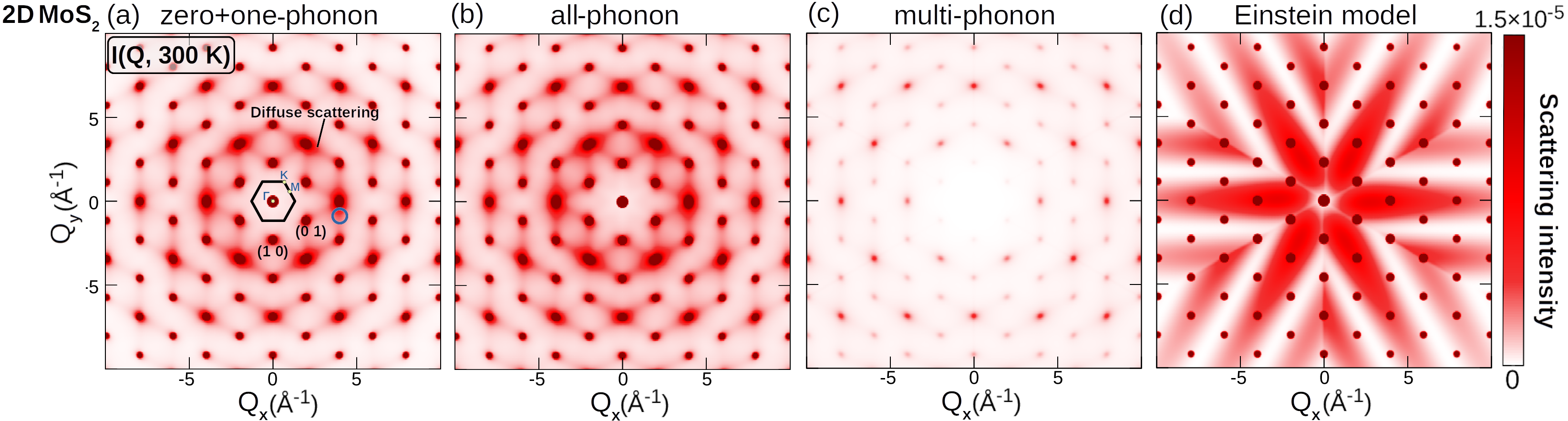}
  \caption{(a) Zero-plus-one-phonon, (b) all-phonon, (c) multi-phonon, and (d) Einstein model scattering intensity 
   of monolayer MoS$_2$ calculated for $T=300$~K.
  In plot (a) we show the fundamental Brillouin zone together with the high-symmetry points $\Gamma$, K and M.
  We also show the (1 0) and (0 1) Bragg peaks. %, as well as an example of diffuse scattering signal. 
  Blue circle indicates a rapid decrease in the diffuse scattering intensity. 
  The sampling of the Brillouin zone was performed using a $50\times50$ {\bf q}-grid. For all plots the scattering 
  intensity is divided by the maximum Bragg intensity, i.e with $I_0(\bQ = {\bf 0},T)$. 
  \label{fig2} 
  }
\end{figure*}

\begin{figure*}[hbt!]
  \includegraphics[width=0.9\textwidth]{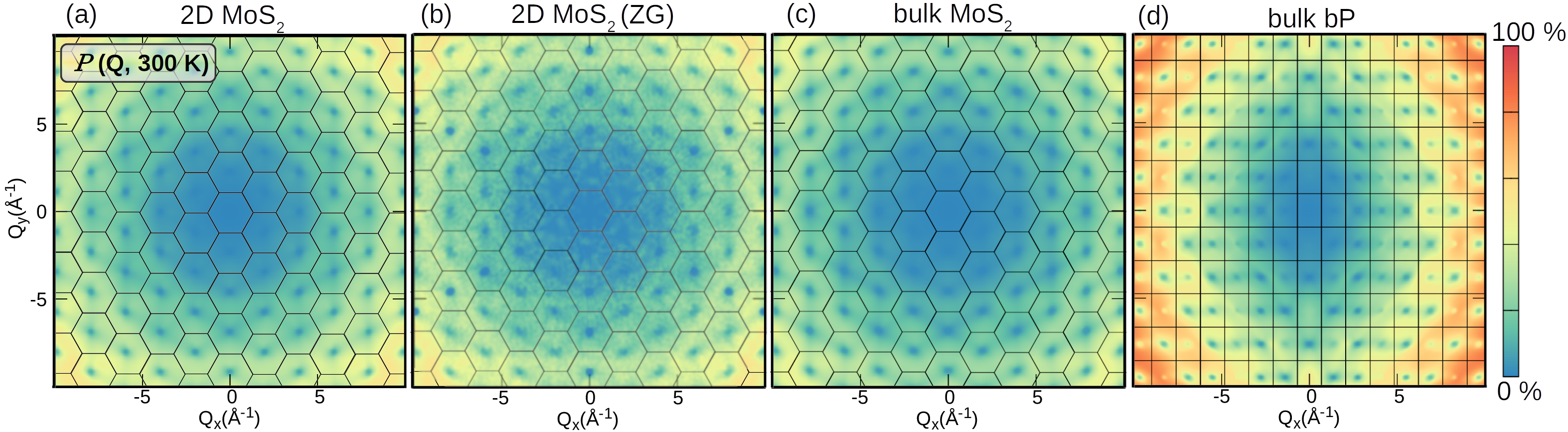}
  \caption{Percentage contribution of multi-phonon interactions to diffuse scattering of (a),(b) monolayer MoS$_2$, 
  (c) bulk MoS$_2$,  and (d) bulk black Phosphorous (bP) calculated as 
  $\mathcal{P} = I_{\rm multi}/(I_{1} + I_{\rm multi}) \times 100$ at $T = 300$~K.
  (a), (c), and (d) represent calculations within the LBJ theory and (b) using ZG displacements. 
  Maps are separated into Brillouin zones to highlight the extent of multi-phonon interactions. 
  \label{fig3} }
\end{figure*}

\subsection{Computational details}  \label{sec.Theory_Methods}

\textit{Ab initio} calculations were performed using planewaves basis sets and  
the PBE generalized gradient approximation~\cite{GGA_Pedrew_1996} to density-functional theory (DFT), 
as implemented in the {\tt Quantum ESPRESSO} software package~\cite{QE,QE_2}.
We used the primitive cells of 2D transition-metal dichalcogenides (MoS$_2$, MoSe$_2$, WSe$_2$, 
and WS$_2$ with space group P${\bar{6}}$m2), CdI$_2$ (P${\bar{3}}$m1),  
graphene (P6/mmm), bulk MoS$_2$ (P6$_3$/mmc), and bP (Cmce) that contain 3, 3, 2, 6, and 4 atoms, respectively.
We employed Goedecker-Hartwigsen-Hutter-Teter norm-conserving pseudopotentials~\cite{GTH_1996,HGH_1998} for 
all monolayers and bulk MoS$_2$, and Troullier-Martins~\cite{Troullier_Martins_1991} 
norm-conserving pseudopotentials for bP. The planewaves kinetic energy cutoff was set 
to 80~Ry for graphene, 90~Ry for bP, 100~Ry for CdI$_2$, 120~Ry for MoS$_2$,
130~Ry for MoSe$_2$, WSe$_2$, and WS$_2$. 
Self-consistent-field calculations were performed using Brillouin zone ${\bf k}$-grids 
of $10\!\times\!10\!\times\!1$ (monolayers MoS$_2$ MoSe$_2$, WSe$_2$, WS$_2$ and graphene), 
$14\!\times\!14\!\times\!1$ (monolayer CdI$_2$), 
$10\!\times\!10\!\times\!3$ (bulk MoS$_2$), and $12\!\times\!10\!\times\!10$ (bP) points. 
To avoid interactions between periodic replicas of the monolayers we used an interlayer 
vacuum larger than 15 \AA\, and a truncated Coulomb interaction~\cite{Sohier_2017}.
The optimized lattice parameters for monolayers are $a = 3.17$~\AA\, (MoS$_2$), $3.32$~\AA\,(MoSe$_2$), 
$3.31$~\AA\,(WSe$_2$), $3.18$~\AA\,(WS$_2$), $2.47$~\AA\,(graphene), $4.33$~\AA\,(CdI$_2$); 
$a = 3.191$~\AA \, and $c= 12.43$~\AA\, for bulk MoS$_2$;
$a =3.307$~\AA, $b= 4.554$~\AA, and $c =11.256$~\AA\, for bP.
We determined the interatomic force constants by means of density-functional perturbation theory~\cite{Baroni_2001}  
using Brillouin zone ${\bf q}$-grids of $8\!\times\!8\!\times\!1$ (monolayers), $8\!\times\!8\!\times\!2$ (bulk MoS$_2$), 
and $5\!\times\!5\!\times\!5$ (bP) points.

The zero ($I_{\rm 0}$), one ($I_{\rm 1}$), and all ($I_{\rm all}$) phonon 
scattering intensities were calculated employing Eqs.~\eqref{eqa1.12_b},~\eqref{eqa1.13} and~\eqref{eqa1.7}, respectively.  
%, $I_{\rm 0}(\bQ,T) + I_{\rm 1}(\bQ,T)$,
For the calculation of the exponent of the Debye-Waller [Eq.~\eqref{eqa1.8}] and phononic [Eq.~\eqref{eqa1.8_b}] factors,
the full sets of phonon eigenmodes and eigenfrequencies were obtained by using standard Fourier interpolation of 
dynamical matrices on ${\bf q}$-grids of $50\!\times\!50\!\times\!1$ (monolayers) 
and $50\!\times\!50\!\times\!50$ (bulk systems) points,
unless specified otherwise. $\bQ$-grids of the same size were employed to sample 
the scattering pattern per Brillouin zone of each system.
We must emphasize that it is erroneous to compute the all-phonon scattering intensity using  
$\bQ$- and ${\bf q}$-grids of different density, since this violates the momentum 
selection rule and gives rise to artefacts in the phonon-diffuse pattern. 
For MoS$_2$ systems we show patterns calculated in the $Q_x$-$Q_y$ planes 
at $Q_z = 0$, where $Q_x$, $Q_y$, and $Q_z$ are the Cartesian components of $\bQ$. 
bP patterns are obtained as the average of the scattering intensities 
at $Q_z = 0$ and $Q_z = 2 \pi / c = 0.56$~\AA$^{-1}$ planes.
Simulating the zero-order Laue zone ($Q_z = 0$ plane) and the 
first-order Laue zone ($Q_z = 2 \pi/c$ plane) reproduces more Bragg peaks 
observed in the experiment which we attribute to stacking faults in the sample~\cite{Jesson1990,ReyesGasga2008,Gomez2014}.
The atomic scattering amplitudes $f_\k(\bQ)$ for each atom were obtained analytically as a sum of 
Gaussians~\cite{Vand_1957} using the parameters in Ref.~[\onlinecite{Peng_book}].
For the calculation of the full maps of hexagonal (monolayers and bulk MoS$_2$) 
and orthorhombic (bP) systems, we applied a six-fold and 
four-fold rotation symmetry around the $\Gamma$-point.

The set of special displacements [Eq.~\eqref{eq.realdtau_method00}] were generated via the 
{\tt ZG} executable ({\tt ZG.x}) of the {\tt EPW} software package~\cite{Ponce_2016_EPW}. 
The general procedure for applying SDM is described in Ref.~[\onlinecite{Zacharias_2020}]. 
In short, here we 
(i) used the same ${\bf q}$-grid as for the Debye-Waller and phononic factors,
(ii) ordered the phonon eigenmodes and frequencies 
along a simple space-filling curve that passes through all ${\bq}$-points,
(iii) ensured similarity by enforcing a smooth Berry connection between the phonon eigenmodes 
at adjacent ${\bq}$-points, and (iv) assigned $2^{n-1}$ unique combinations of $n$ 
signs $\{S_{\bq  \nu},\cdots, S_{\bq  \nu'}\}$ to every $2^{n-1}$ ${\bq}$-points, where $n$, here,
 is equal to the number of phonon branches. These choices together with the dense 
grids employed guarantee fast minimization of Eq.~\eqref{eq.minimiz_function}. 
The ZG scattering intensity was calculated with Eq.~\eqref{eq.ZG_struct_factor} 
using the same $\bQ$-grid as for the LBJ scattering intensity. 
Notably, implementing Eq.~\eqref{eq.ZG_struct_factor} is much more straightforward
than Eq.~\eqref{eqa1.7}. Hence, SDM serves as a guide for validating
our calculations of the LBJ diffuse scattering intensity. 

The code ({\tt disca.x}) used for the calculation of all phonon contributions 
to diffuse scattering is available at the {\tt EPW/ZG} tree. 
The ZG scattering intensity was computed with {\tt ZG.x}.
It is worth noting that the fine grids employed for the purposes of this work do not have 
high computational requirements since they do not involve extra DFT steps.
In fact, these codes act as post-processing steps and allow for the rapid evaluation of 
the (ZG or LBJ) scattering intensity of any material, provided that the interatomic force constants 
have already been computed. No restrictions are imposed on the methodology followed for the evaluation of 
interatomic force constants; this can be by means, for example, of density-functional perturbation 
theory~\cite{Baroni_2001}, the frozen-phonon method~\cite{Ackland_1997}, 
the self-consistent harmonic approximation~\cite{Errea_2014}, or {\it ab initio} molecular dynamics~\cite{Hellman_2011}. 

\begin{figure}[ht!]
  \includegraphics[width=0.44\textwidth]{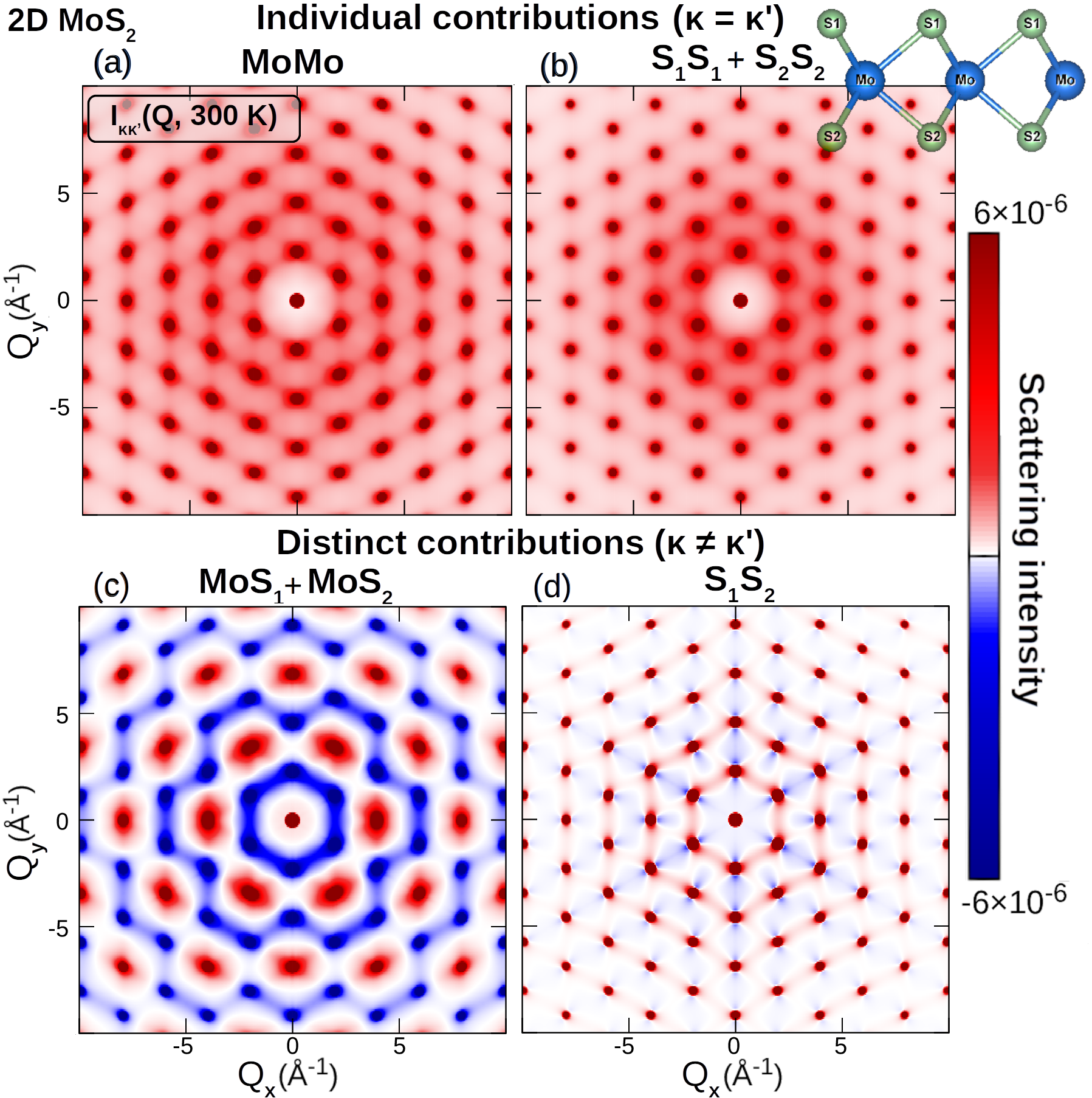}
  \caption{Individual and distinct atomic contributions to the all-phonon scattering intensity 
   of monolayer MoS$_2$ calculated for $T=300$~K. (a) and (b) is for the Mo and S individual scattering terms.
  (c) and (d) is for the MoS and S$_1$S$_2$ distinct scattering terms. 
  The Brillouin zone sampling was performed using a $50\times50$ {\bf q}-grid. 
  Data is divided by the Bragg intensity at the centre of the Brillouin zone, 
  i.e with $I_0(\bQ = {\bf 0},T)$. \label{fig4} }
\end{figure}

\begin{figure*}[t!]
  \includegraphics[width=0.99\textwidth]{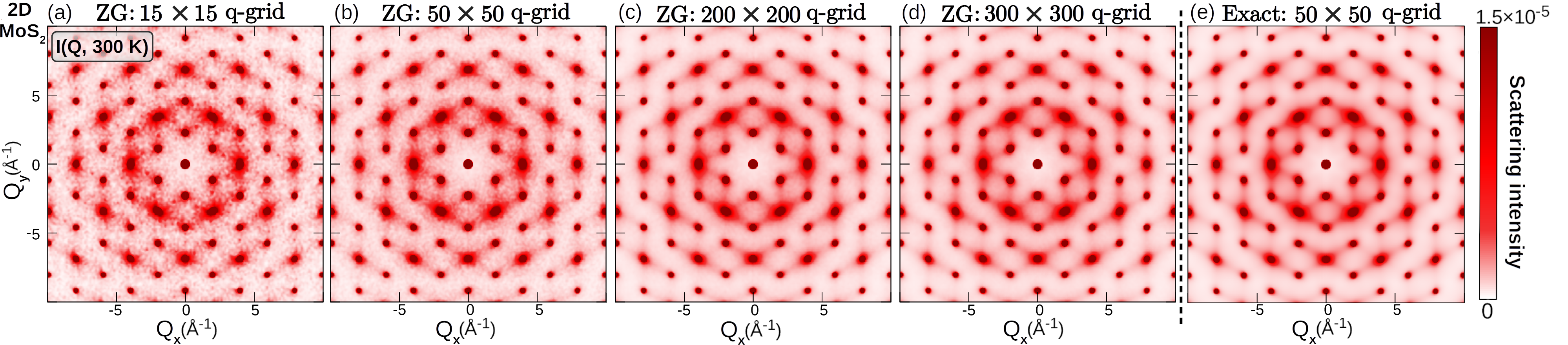}
  \caption{(a)-(d) Convergence of the ZG scattering intensity of monolayer MoS$_2$ at $T=300$~K 
  with respect to the Brillouin zone sampling. (e) Exact all-phonon scattering intensity 
  calculated using Eq.~\eqref{eqa1.7}. All data is divided by the Bragg peak at the centre 
  of the Brillouin zone, i.e with $I_0(\bQ = {\bf 0}, T)$. \label{fig5} }
\end{figure*}

\section{Results} \label{sec.results}

\subsection{2D MoS$_2$}\label{sec.res_2D_MoS2}

Figures~\ref{fig2}(a), (b), and (c) show the zero-plus-one-phonon, multi-phonon, and all-phonon 
scattering intensities at $T=300$~K in the reciprocal space of monolayer MoS$_2$. 
All-phonon and zero-plus-one-phonon excitations were accounted for via Eq.~\eqref{eqa1.7} and 
combining Eqs.~\eqref{eqa1.8}~and~\eqref{eqa1.8_b}, respectively;  
full computational details are provided in Sec.~\ref{sec.Theory_Methods}.
Both sets of data have been normalized such that the scattering intensity at the zone-center is equal to 1. 
The multi-phonon scattering intensity was obtained from $I_{\rm multi} = I_{\rm all} - I_{\rm 0} - I_{\rm 1}$. 
Our results show that the diffuse pattern of monolayer MoS$_2$ is determined to a large extent  
by one-phonon scattering, while multi-phonon interactions play a secondary role without introducing 
new features. % in the diffraction pattern.
To quantitatively assess the effect of multi-phonon processes on the diffuse pattern we report in Fig.~\ref{fig3}(a) 
the percentage $ \mathcal{P} = I_{\rm multi}/(I_{1} + I_{\rm multi}) \times 100$
as a function of $\bQ$.%, i.e. the distance of the scattering vector from the zone-center.  
The response of the scattering intensity to multi-phonon excitations increases 
as we move radially outwards from the center, exceeding 50\% for $|\bQ| \geq 12$\,\AA$^{-1}$. 
However, when $\bQ \sim \bG$ (centers of Brillouin zones), we find that single-phonon contributions dominate
and $ \mathcal{P}$ reduces significantly.

\begin{figure*}[htb!]
\includegraphics[width=0.85\textwidth]{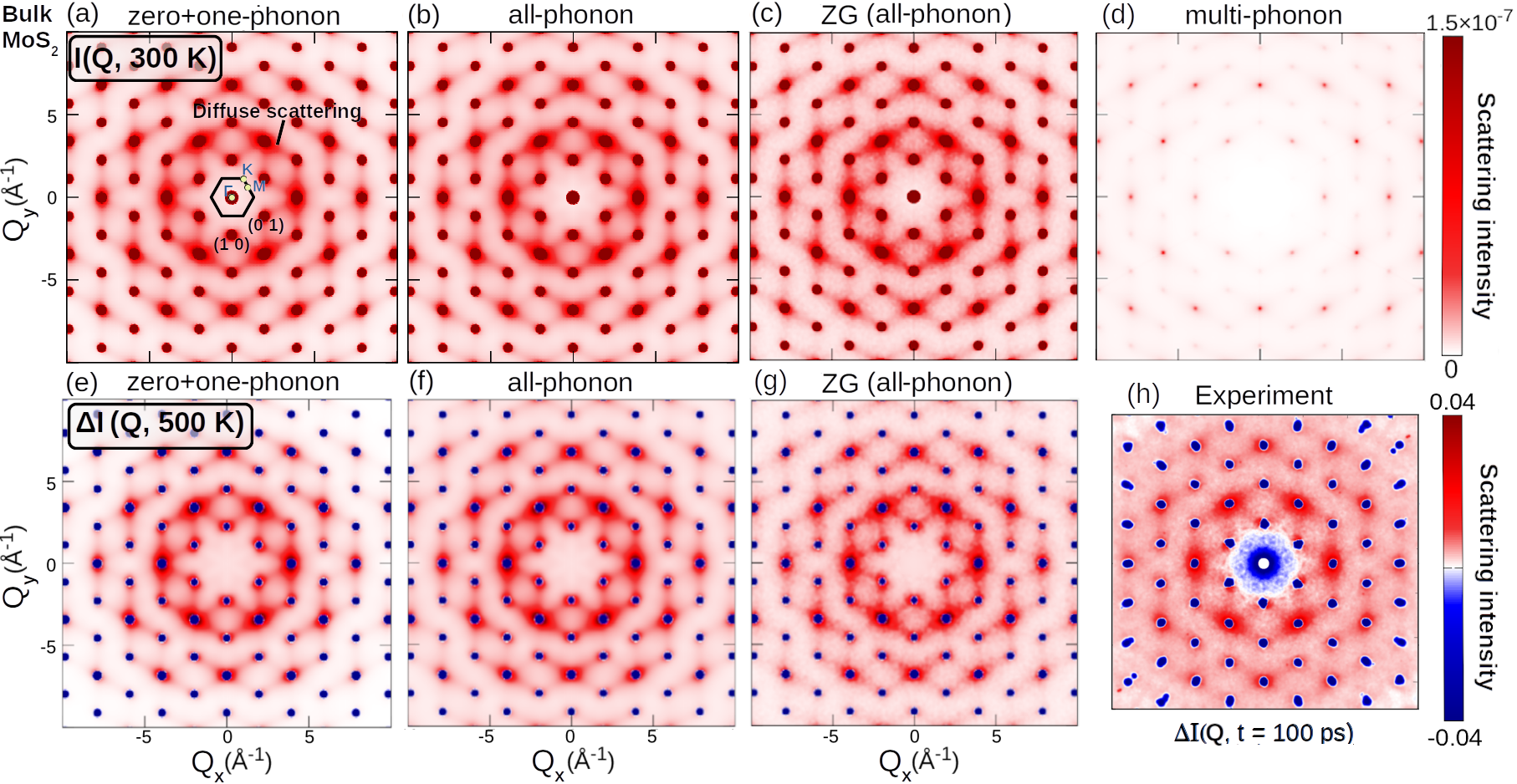}
  \caption{(a) Zero-plus-one-phonon, (b) all-phonon, (c) ZG (all-phonon),  and (d) multi-phonon scattering intensity 
  of bulk MoS$_2$ calculated for $T=300$~K. The calculated intensities are divided by the Bragg intensity 
  at the centre of the Brillouin zone, i.e with $I_0(\bQ = {\bf 0},T)$.
  In plot (a) we show the fundamental Brillouin zone together with the high-symmetry points $\Gamma$, K, and M.
  We also show the (1 0) and (0 1) Bragg peaks.  
  In plot (a) we indicate the (1 0) and (0 1) Bragg peaks, as well as regions
  of diffuse and Bragg scattering. (e) Zero-plus-one-phonon, (f) all-phonon and (g) ZG (all-phonon) diffuse maps
  of bulk MoS$_2$ calculated as $\Delta I (\bQ,500 \,{\rm K}) = I(\bQ,500 \,{\rm K}) - I(\bQ, 300\, {\rm K})$,
  corresponding to the temperature difference estimated from the experiments.
  (h) Experimental scattering signals of bulk MoS$_2$ measured at 100~ps. Signals are divided by the maximum count
  due to elastic scattering. For comparison purposes, the simulated data is multiplied by $500000$ to match the 
  experiment. The sampling of the Brillouin zone was performed using a $50\times50\times50$ {\bf q}-grid.
  \label{fig6} 
  }
\end{figure*}

\begin{figure*}[hbt!]
\includegraphics[width=0.85\textwidth]{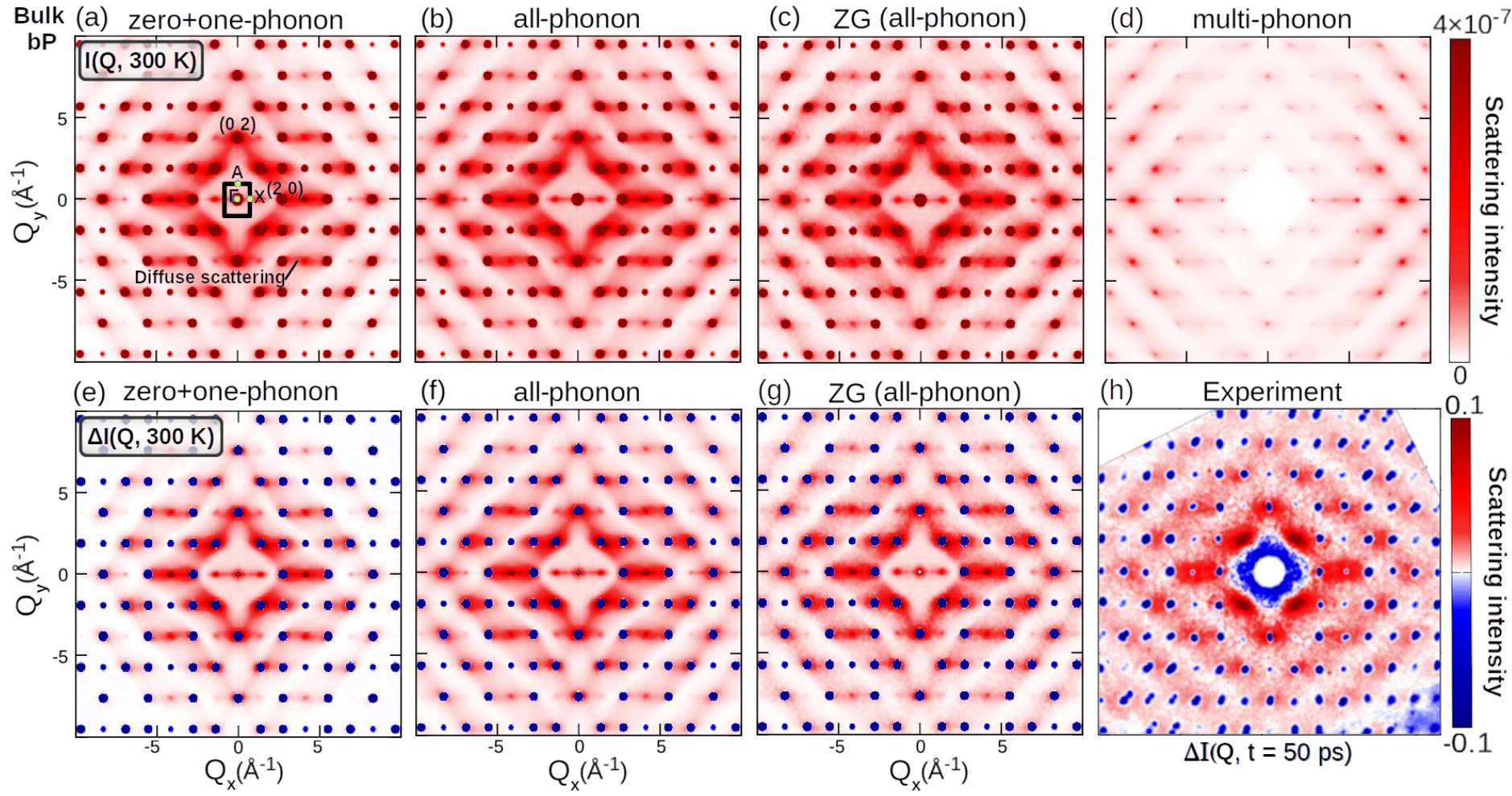}
  \caption{
  (a) Zero-plus-one-phonon, (b) all-phonon, (c) ZG (all-phonon),  and (d) multi-phonon scattering intensity 
  of bulk black Phosphorous (bP) calculated for $T=300$~K. The calculated intensities are divided by the 
  elastic scattering at the central Bragg peak, $I_0(\bQ = {\bf 0},T)$. In plot (a) we show the fundamental 
  Brillouin zone together with the high-symmetry points $\Gamma$, A and X. We also show the (2 0) and (0 2) 
  Bragg peaks. %, as well as an example of diffuse scattering signal. 
  (e) Zero-plus-one-phonon, (f) all-phonon, and (g) ZG (all-phonon) difference scattering maps
   of bulk black Phosphorous calculated as 
  $\Delta I (\bQ,300 \,{\rm K}) = I(\bQ,300 \,{\rm K}) - I(\bQ, 100\, {\rm K})$,
   compatible with experimental conditions. 
  (h) Experimental difference scattering signals measured at 50 ps. 
  Signals are divided by the maximum count due to elastic scattering. Simulations are multiplied by $400000$ 
  to match the experimental maximum intensity~\cite{arXiv:2103.10108}.  
  The sampling of the Brillouin zone was performed using a $50\times50\times50$ {\bf q}-grid.
  \label{fig7} 
  }
\end{figure*}

\begin{figure}[htb!]
  \includegraphics[width=0.47\textwidth]{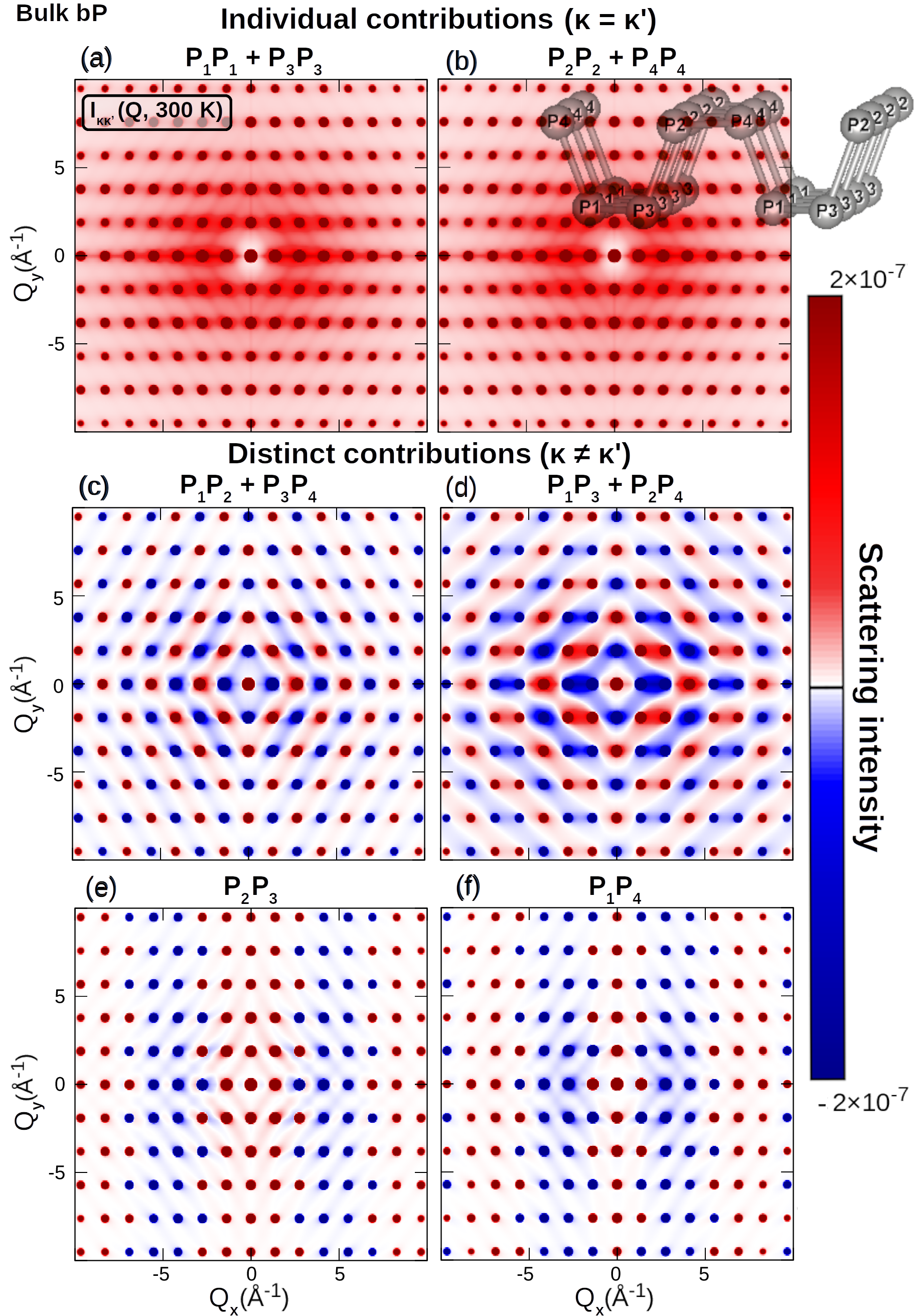}
  \caption{Individual and distinct atomic contributions to the all-phonon scattering intensity 
  of bulk black Phosphorous calculated for $T=300$~K. (a) and (b) is for the individual P$_1$, P$_2$, P$_3$,
  and P$_4$ contributions. (c), (d), (e) and (f) is for the distinct (and inequivalent) P$_i$P$_j$ contributions.
  We also report a ball and stick model of bP. The Brillouin zone sampling was performed using 
  a $50\times50\times50$ {\bf q}-grid and data is divided by the Bragg intensity at the centre 
  of the Brillouin zone, i.e with $I_0(\bQ = {\bf 0},T)$. \label{fig8} }
\end{figure}

\begin{figure*}[hbt!]
  \includegraphics[width=0.75\textwidth]{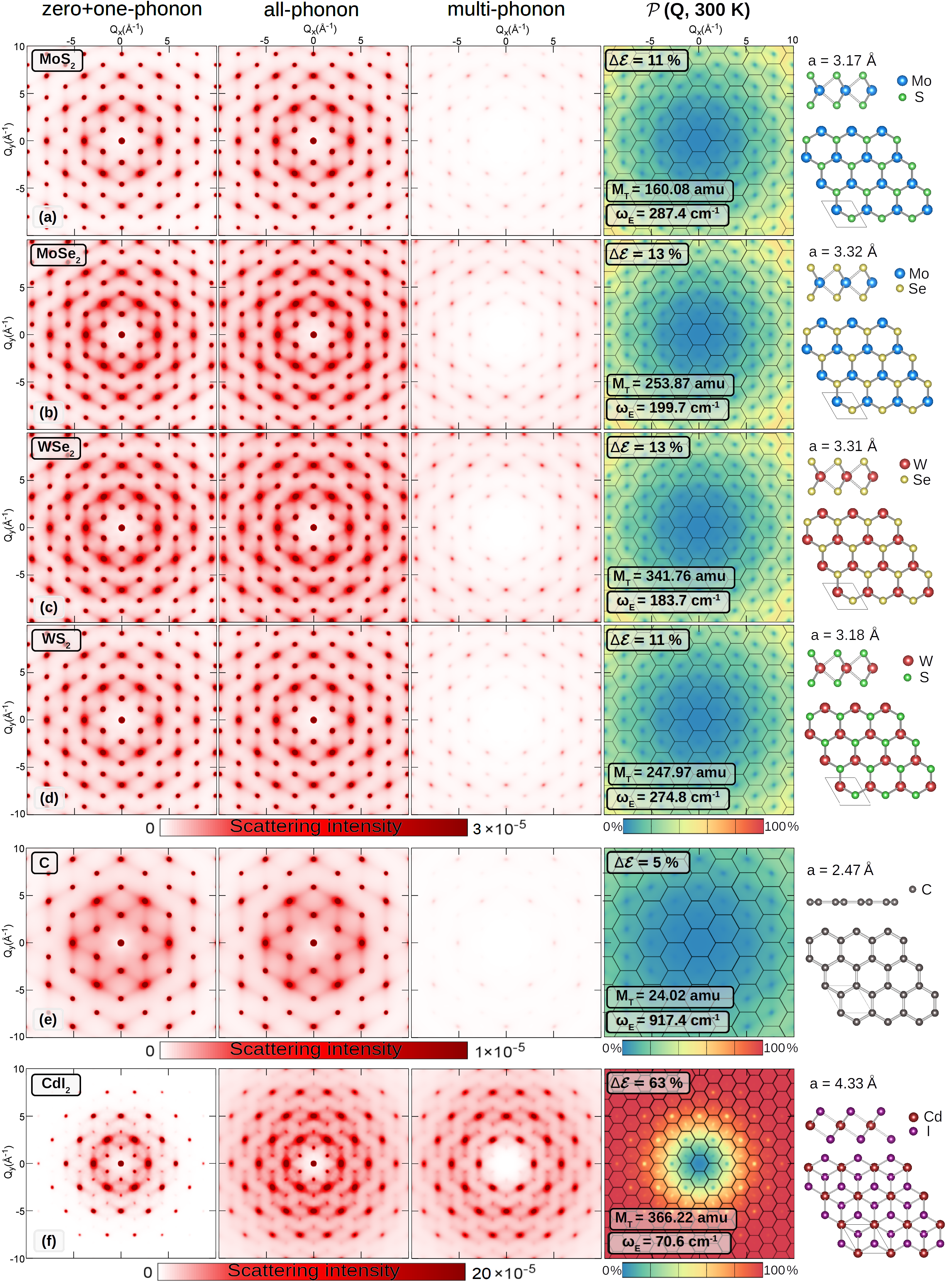}
  \caption{
  Zero-plus-one-phonon, all-phonon, and multi-phonon scattering intensities, and percentage contribution of
  multi-phonon processes  to diffuse scattering intensity $\mathcal{P}$ calculated for 2D (a) MoS$_2$, 
  (b) MoSe$_2$, (c) WSe$_2$, (d)  WS$_2$, (e) graphene, and (f) CdI$_2$ all at $T=300$~K.
  The energy transfer to the crystal from multi-phonon scattering $\Delta \mathcal{E}$, 
  the total atomic mass per unit-cell in atomic mass units (amu) $M_T$, and the mean 
  phonon frequency $\omega_{\rm E}$ are indicated on each plot. The sampling of the Brillouin 
  zone was performed using a $50\times50$ {\bf q}-grid  and data is divided by the maximum Bragg 
  intensity, i.e with $I_0(\bQ = {\bf 0},T)$. We also provide the ball and stick model, primitive-cell, and 
  optimized lattice parameter $a$ of each structure.  \label{fig9} }
\end{figure*}

\begin{figure}[t!]
 \includegraphics[width=0.44\textwidth]{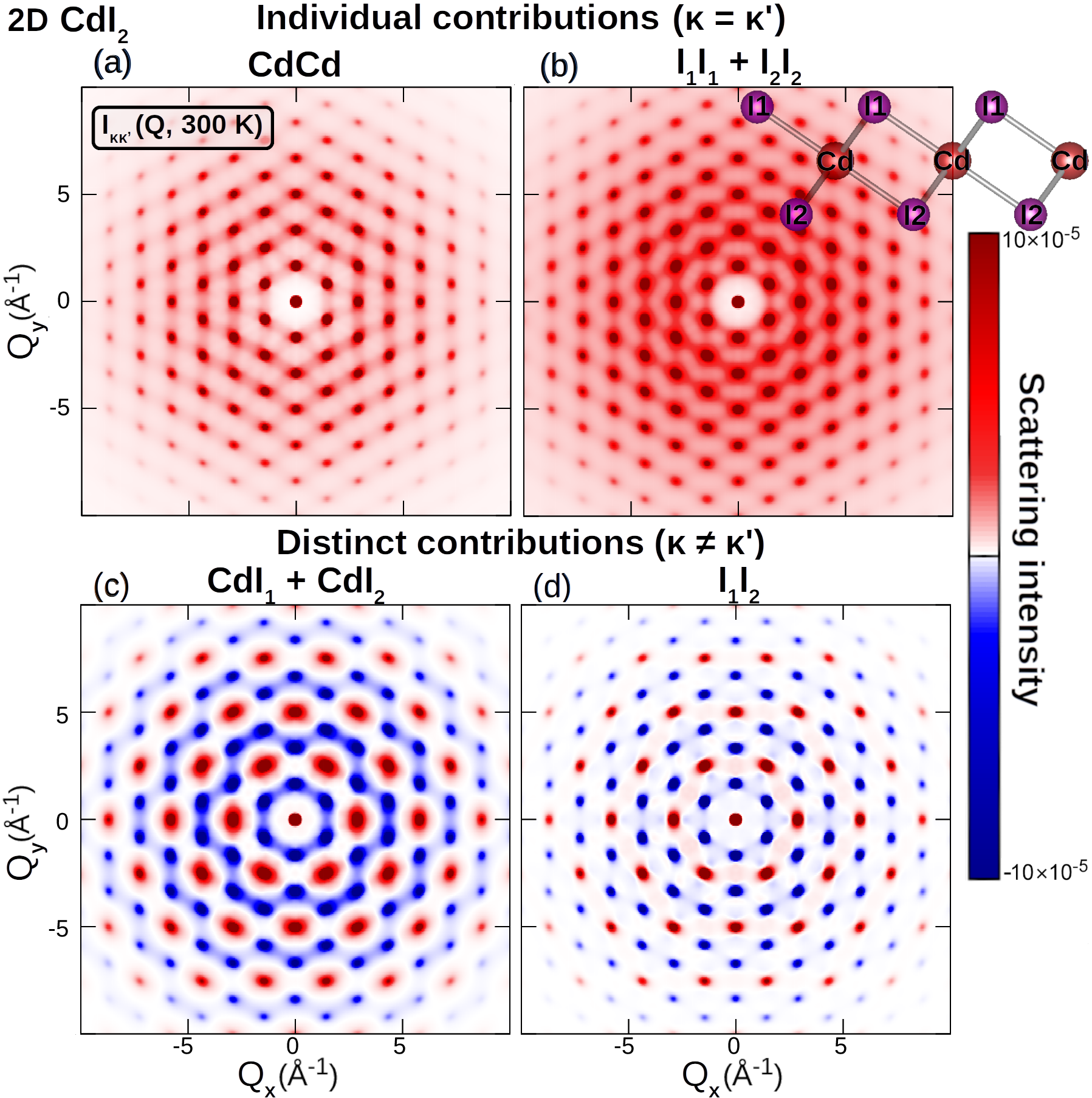}
  \caption{
  Individual and distinct atomic contributions to the all-phonon scattering intensity 
   of monolayer CdI$_2$ calculated for $T=300$~K. (a) and (b) is for the Cd and I individual
  contributions. (c) and (d) is for the CdI and I$_1$I$_2$ distinct terms. 
  The Brillouin zone was sampled using a $50\times50$ {\bf q}-grid. 
  Data is divided by the maximum Bragg intensity, 
  i.e with $I_0(\bQ = {\bf 0},T)$. \label{fig10} }
\end{figure}

In Fig.~\ref{fig2}(d), we present the total scattering intensity in the Einstein model calculated using
Eq.~\eqref{eqa1.7_Einstein} and setting $\omega_{\rm E} = 287.4$ cm$^{-1}$. With no surprise, the Einstein model
% Bulk \omega_{\rm E} 285.66
fails completely to explain diffuse scattering in 2D MoS$_2$ resembling scattering
patterns calculated for isotropic systems~\cite{Muller_2001}.
However, this approximation can provide a rough prediction of the multi-phonon contributions to 
diffuse scattering by evaluating the total energy transfer
 to the crystal, $\Delta \mathcal{E}$, as defined in the parallel paper, Ref.~[\onlinecite{arXiv:2103.10108}]. 
For the range presented in Fig.~\ref{fig2}, the Einstein model yields $\Delta \mathcal{E}_{\rm E} = 10$\% 
in very close agreement with the exact value $\Delta \mathcal{E} = 11$\% obtained within the LBJ theory. 
It is worth noting that a corresponding calculation of the percentage $\mathcal{P}$ will miss 
the reduced contribution of multi-phonon interactions at the Bragg peaks~\cite{arXiv:2103.10108}.

To understand the main features in the scattering pattern of monolayer MoS$_2$ we examine 
the individual atomic ($\k = \k'$) and interatomic ($\k \neq \k'$) terms entering Eq.~\eqref{eqa1.13}. 
Figures~\ref{fig4}(a) and (b) show our calculations 
for the Mo and S individual contributions to the all-phonon scattering intensity. In both cases, the
Bragg scattering amplitude decreases gradually with the distance from the zone-center. In view of Eq.~\eqref{eqa1.12_b}, 
this gradual decrease is attributed solely to the attenuation coming from the Debye-Waller and atomic form factors, since 
the modulation factor  $\text{cos}\big[ \bQ \cdot (\bt_\k - \bt_{\k'})\big]$ simplifies to $1$ for the individual terms. 
The same holds for the strong diffuse scattering concentrated in the vicinity of the Bragg peaks. 
Within the first Brillouin zone, the patterns exhibit a relatively weak intensity as a result of the small transferred momenta.

Figures~\ref{fig4}(c) and (d) show the response of the all-phonon scattering intensity
to each inequivalent distinct pairing: MoS and S$_1$S$_2$.
It is evident that MoS collective displacements tend to decrease, or increase,
the Bragg scattering intensity depending on the factor $\text{cos}\big[ \bQ \cdot (\bt_\k - \bt_{\k'})\big]$ 
and the symmetry of the structure. In particular, our analysis shows that for a Bragg scattering vector $\bQ = (h \, k)$,
the MoS pairs enhance (suppress) the total intensity when $|h-k| = 3n$ ($\neq 3n$), where $h$, $k$ and $n$ are integers.
MoS paired thermal fluctuations also contribute to the diffuse scattering constructively, or destructively, explaining the 
rapid decrease in the scattering probability between adjacent Bragg peaks, 
as indicated by the blue circle in Fig.~\ref{fig2}(a).
For S$_1$S$_2$ distinct terms, the cosine modulation factor simplifies to $1$ owing to the 
trigonal prismatic coordination of the S atoms, thereby enhancing Bragg scattering. 
The correlated vibrational motion between sulphide atoms tends 
to reduce phonon-induced scattering in a way that the intensity of the
star-like domain formed within the first and second order Brillouin zones 
of monolayer MoS$_2$ is enhanced.

\subsection*{Evaluation of the all-phonon scattering intensity using the ZG displacement}

As described in Secs.~\ref{ZG_theory}~and~\ref{sec.Theory_Methods}, SDM constitutes an alternative way for the evaluation 
of the scattering intensity and can be used as a tool to further verify our implementation of Eq.~\eqref{eqa1.7}.  
Here we provide a detailed convergence test, using the example of monolayer MoS$_2$, and demonstrate that 
the two approaches give identical results in the limit of dense Brillouin zone sampling. 

In order to analyze the convergence behavior of the SDM, 
in Figs.~\ref{fig5}(a)-(d) we plot the dependence of the ZG scattering intensity on the 
$\bq$-grid used to generate special displacements.
For comparison purposes, in Fig.~\ref{fig5}(e) we also present the data obtained 
using the exact expression in Eq.~\eqref{eqa1.7}. The ZG scattering intensity calculated 
for a $15\times15$ $\bq$-grid, commensurate with the supercell size of realistic ZG 
DFT-calculations, compares well with the exact result and reveals all main features in the 
patterns. Deviations from the Bragg and inelastic scattering 
appear as a statistical background noise and are explained by the error in the evaluation 
of the ZG observable. We remark that calculations of the difference images 
between ZG and exact patterns show that discrepancies are more prominent at the Bragg peaks, 
as a result of the two extra terms entering the function $\Delta_{\k\k'}(\bQ,T)$ when $\bQ = \bG$  
[Eq.~\eqref{ZG_sf_5} of the Appendix]. As shown in Figs.~\ref{fig5}(b)-(d), the error is 
alleviated by using finer $\bq$-grids and vanishes in the limit of dense Brillouin sampling, 
i.e. for a $300\times300$ $\bq$-grid. The agreement between the two methods is further 
substantiated in Fig.~\ref{figA1}, where the multi-phonon contribution to the all-phonon 
scattering intensity is identical when calculated with ZG displacements, or with Eq.~\eqref{eqa1.7}. 
A similar conclusion can be drawn by comparing $\mathcal{P}$ in Figs.~\ref{fig3}(a) and~(b). This successful
comparison provides the first rigorous numerical proof that SDM can seamlessly capture higher-order terms in 
the Taylor expansion of the observable.    

Following the above analysis, it becomes apparent that ZG displacements lead precisely to 
the thermally distorted structure that reproduces the all-phonon diffuse scattering.
Although thermal diffuse scattering is fundamentally related to the phonon properties of the crystal, this concept 
reinforces the use of ZG displacements for the evaluation of temperature-dependent electronic and optical properties 
of solids, as attested in Refs.~\cite{Zacharias_2016,Tathagata_2017,Gunst_2017,Zhang_2018,Kang_2018,Karsai_2018,Karsai_2018_b,
Palsgaard_2018,Huang_2019,Novko_2019,Zacharias_2020,Zhang_2020,Ha_Viet_Anh_2020,Alexey_2020,Zacharias_PK_2020,Liu_2021_arxiV,Huang2021}.
It is also evident that ZG calculations can capture accurately all terms in the Taylor expansion of the observable of interest, 
and thus can serve as a tool for the assessment of multi-phonon effects, including carrier-multi-phonon coupling.
On top of that, SDM can be upgraded straightforwardly for the calculation of ultrafast phonon-diffuse
data~\cite{Seiler2021} and other non-equilibrium electron-phonon mediated properties. 
In particular, non-equilibrium phonon occupations computed by the Boltzmann 
transport equation~\cite{Caruso_2021} can enter directly Eq.~\eqref{eq_SDM_3}, and hence allow 
for the generation of time-resolved ZG displacements via Eq.~\eqref{eq.realdtau_method00}. 
This will, in turn, significantly simplify the interpretation of ultrafast 
phenomena, providing a physical picture with respect to real-space displacements.

\subsection{Bulk MoS$_2$} \label{sec.res_bulk_MoS2}

Figures~\ref{fig6}(a)-(c) show the zero-plus-one phonon, all-phonon, and ZG
scattering patterns of bulk MoS$_2$ at $T=300$~K. All sets of data have been normalized such that 
the intensity at the zone-center is equal to 1. 
The scattering pattern of bulk MoS$_2$ is qualitatively identical to the one of
its monolayer counterpart shown in Fig.~\ref{fig2}. 
Quantitatively, the major difference is that the intensity of Bragg
scattering in bulk MoS$_2$ is about two orders of magnitude higher. 
These findings suggest that collective displacements between any two atoms that lie in separate
MoS$_2$ layers do not participate actively in diffuse scattering. Indeed, 
our analysis (not shown) confirms that these distinct pairs contribute predominantly to Bragg scattering 
and very little to diffuse scattering. 
Similarly to the monolayer MoS$_2$, the main characteristics in the diffuse pattern arise from 
the MoS correlated displacements. %, while the star-like domain is due to the S$_1$S$_2$ pair. 

In Fig.~\ref{fig6}(d) we present the multi-phonon structure factor map
of bulk MoS$_2$, obtained as the difference between 
the all-phonon and zero-plus-one-phonon diffuse patterns, i.e. $I_{\rm multi} = I_{\rm all} - I_{0} - I_{1}$.
Our results reveal that scattering beyond one phonon does not smear out the fundamental information enhancing slightly 
the scattering signal.
This observation is further supported by Fig.~\ref{fig3}(c), which shows that 
the multi-phonon contribution to inelastic scattering, $\mathcal{P}$, never dominates over one-phonon processes 
for any $|\bQ| \leq 14$ \AA$^{-1}$.

In Figs.~\ref{fig6}(e)-(h) we compare the zero-plus-one-phonon, all-phonon, and ZG difference scattering patterns of
bulk MoS$_2$ with the experimental signals measured at a pump-probe delay of 100 ps, 
$\Delta I(\bQ, t = 100 \, {\rm ps})$. At this time delay, we assume that phonon 
thermalization is reached~\cite{Seiler2021}.
Blue and red colouring represent a decrease and an increase in the relative scattering intensity, respectively. 
Bragg peaks appear as blue dots since the exponent of the Debye-Waller factor, $-W_{\k} (\bQ,T)$, 
is reduced with increasing temperature. The agreement between theory and experiment is excellent, except that we 
underestimate the background diffuse scattering. This discrepancy is diminished when multi-phonon 
interactions via Eq.~\eqref{eqa1.7}, or ZG displacements, are accounted for. 
Despite multi-phonon scattering, the background observed experimentally can be due to many others factors, 
such as multiple scattering events and inelastic scattering on plasmons~\cite{Zhong_Lin_Wang,Sascha_2011,Zahn_2020}.

\subsection{Bulk black phosphorus} \label{sec.res_bulk_bP}

Figures~\ref{fig7}(a) and (b) show the scattering patterns of bulk bP at $T=300$~K calculated using 
the zero-plus-one-phonon and all-phonon expressions, respectively. For completeness, we also 
report the ZG scattering intensity at the same temperature in Fig.~\ref{fig7}(c). 
In Fig.~\ref{fig7}(d), we show the multi-phonon scattering pattern
of bulk bP. Unlike 2D and bulk MoS$_2$, multi-phonon processes in bP strongly enhance diffuse scattering
away from the zone-center revealing, essentially, new diamond-like patterns.  
In Fig.~\ref{fig3}(d), we also disclose the percentage contribution of 
multi-phonon excitations to diffuse scattering intensity, $\mathcal{P}$. 
We find that higher-order processes play the primary role to diffuse scattering for $|\bQ| \geq 8$ \AA$^{-1}$ 
reaching a maximum of 83\% at $|\bQ| = 13$ \AA$^{-1}$. 
It is also evident from Fig.~\ref{fig3} that $\mathcal{P}$ is much more prominent in bP than in MoS$_2$ crystals. 
Using our toy model developed in the parallel paper, Ref.~[\onlinecite{arXiv:2103.10108}], and observing that
the mean frequencies of the three crystals are similar, we can then attribute this different 
behaviour to the lighter mass of phosphorus.  

For completeness, in Figs.~\ref{fig6}(e)-(h) we reproduce the results of the parallel paper, Ref.~[\onlinecite{arXiv:2103.10108}],  
and compare the zero-plus-one-phonon, all-phonon, and ZG difference scattering patterns of
bulk bP with the experimental thermalized signals measured at a pump-probe delay 
of 50 ps, $\Delta I(\bQ, t = 50 \, {\rm ps})$~\cite{Seiler2021}.
Blue/red areas represent decrease/increase in the relative scattering signal. 
Bragg peaks appear as blue dots as a result of the Debye-Waller effect. 
The zero intensity Bragg peaks, present in both calculations and measurements, are connected with the symmetry 
of the structure and can be explained by analysing the interatomic correlations (see below). 
In the experimental diffraction pattern of bP, however, we observe the presence 
of additional forbidden reflections for $h+k=2n+1$. Such reflections were also observed in previous 
works~\cite{Gomez2014}. They may be caused by stacking faults 
or structural deviations at the surface, as bP is well-known to oxidize rapidly.
These additional reflections do not alter the overall picture.
In fact, the agreement between the all-phonon theory and experiment is striking, confirming 
that multi-phonon excitations change diffuse signals
qualitatively and quantitatively~\cite{arXiv:2103.10108}.
In essence, scattering beyond one-phonon is the main mechanism of the formation of the outer
diamond-like domains. % observed in the experiment.  
These features are also present in the ZG scattering difference pattern, validating once again 
the physical meaning of the ZG distorted structure. 
Given the unprecedent agreement between our both sets of calculated 
all-phonon data and measurements~\cite{arXiv:2103.10108}, we exclude a large redistribution 
of diffuse intensity from lower order into higher order Brillouin zones due to Bragg-Bragg 
and Bragg-diffuse multiple scattering~\cite{Ramsteiner2009}.
Our additional analysis, based on the method described in Ref.~[\onlinecite{Ligges2011}], 
also guarantees that multiple scattering is not a critical issue in our measurements.

In Figs.~\ref{fig8}(a) and (b) we report the all-phonon scattering intensity
coming from the displacements of individual phosphorus atoms. 
The diffuse pattern is mostly structureless and the total 
signal fades out with the distance from the central Bragg peak due to the Debye-Waller 
and atomic form factors. As expected, all Bragg peaks are reproduced since scattered 
waves by individual atoms will undergo constructive interference. 
% at the Bragg peaks Since the cosine modulation factor 

Figures~\ref{fig8}(c)-(f) show the response of the all-phonon 
 scattering intensity to displacements between pairs of P atoms. 
The ball and stick model shows the geometric arrangement of atoms in bP.  
It is evident that electrons scattered 
by the collective motion between atoms that lie in the same basal plane, i.e. P$_1$P$_3$ and P$_2$P$_4$, 
 interfere constructively, or destructively, forming diamond-like domains which 
explain the characteristic diffuse pattern observed in the experiment.    
Regarding other pairs of bP atoms, diffuse scattering is rather insensitive to their collective motion. 
This result demonstrates the potential of diffuse scattering experiments to probe  
microscopic phenomena that occur in specific chemical bonds in solids.

\subsection{2D materials} \label{sec.2D_materials}

In this section, we evaluate
the diffuse scattering patterns of five more 2D materials 
in order to gain further insight into the role of multi-phonon processes and 
demonstrate the high-throughput capability of our method. 

Figure~\ref{fig9} shows the zero-plus-one-phonon, all-phonon, and multi-phonon scattering intensities 
of (a) MoS$_2$, (b) MoSe$_2$, (c) WSe$_2$, (d)  WS$_2$, (e) graphene, and (f) CdI$_2$ all calculated at $T=300$~K
within the LBJ theory. We also show the percentage 
$ \mathcal{P} = I_{\rm multi}/(I_{1} + I_{\rm multi}) \times 100$ to provide
a quantitative hierarchy between one- and multi-phonon processes. 
To support our subsequent analysis we report the calculated energy transfer
from multi-phonon processes $\Delta \mathcal{E}$~\cite{arXiv:2103.10108}, the total 
atomic mass per unit-cell $M_T = \sum_\k M_\k$, and the Einstein phonon frequency 
$\omega_{\rm E}$ of each 2D material. At this point, we recall that $\Delta \mathcal{E}$ scales 
inversely proportional with the atomic masses and phonon frequencies of the system. 
All transition-metal dichalcogenides (MoS$_2$, MoSe$_2$, WSe$_2$, and  WS$_2$) share 
the same space group (P$\bar{6}$m2) and exhibit similar diffraction and phonon-diffuse patterns.  
Importantly, when the sulfide atoms are replaced by the heavier selenium 
in WX$_2$ and MoX$_2$ (X indicates the chalcogen atom), we obtain a subtle enhancement 
of the phonon-induced scattering intensities and $\Delta \mathcal{E}$ by 2\%. 
In both cases, $M_T$ becomes larger by more than 35\%, but $\omega_{\rm E}$ is reduced 
by $\sim 33$\% indicating that the change in the phonon frequencies is the primary 
measure for estimating the extent of multi-phonon processes. 

This conclusion can be further justified by our analysis 
for graphene, shown in Fig.~\ref{fig9}(e). Our results reveal almost identical 
one-phonon and all-phonon diffuse patterns, as well as a small percentage 
$\mathcal{P}$ across the reciprocal space. We find the energy 
transfer due to multi-phonon excitations $\Delta \mathcal{E}$ to be as low as $5$\%. 
Our value comes as no surprise, despite the light mass of carbon atoms. 
In particular, the small $\Delta \mathcal{E}$ is driven by the relatively large 
Einstein phonon frequency ($\omega_{\rm E} = 917.4$~cm$^{-1}$) of graphene, being 
3--4 times larger than $\omega_{\rm E}$ reported for transition-metal dichalcogenides.
It is also evident that employing the one-phonon structure factor is an accurate and 
reliable practise for investigating diffuse scattering signals in materials exhibiting 
large mean phonon frequencies~\cite{Cotret_2019}.

At variance with graphene, multi-phonon processes make a prominent impact on 
the diffuse pattern of 2D CdI$_2$ (space group P$\bar{3}$m1), as shown in Fig.~\ref{fig9}(f). 
We can readily see that the one-phonon scattering theory reveals negligible diffuse scattering for
$|\bQ| > 5$~\AA$^{-1}$, missing important features of the all-phonon scattering map. 
It is also striking that multi-phonon excitations dominate inelastic scattering beyond
the fundamental Brillouin zone, giving $\Delta \mathcal{E} = 63$\%. Based on our previous
discussion, this value is consistent with the relatively small mean phonon frequency of 
CdI$_2$, $\omega_{\rm E} = 70.6$~cm$^{-1}$. The value of $\omega_{\rm E}$
also explains the rapid Debye-Waller damping of the Bragg intensities 
at large scattering wavevectors [see Eq.~\eqref{eqa1.8}]. Unlike the diffraction patterns 
of transition-metal dichalcogenides, we observe that for a Bragg scattering 
vector $\bQ = (h \, k)$, the total intensity is reduced when $|h-k| \neq 3n$. 
%where $h$, $k$ and $n$ are integers. 
To shed light on this result, we report the individual and distinct atomic contributions 
to the all-phonon scattering pattern of monolayer CdI$_2$, as shown in Fig.~\ref{fig10}. 
Apart from the pronounced Debye-Waller damping in CdI$_2$, the main difference between 
the diffraction patterns of MoS$_2$ and CdI$_2$ are due to the distinct contributions 
from S$_1$S$_2$ [Fig.~\ref{fig4}(d)] and I$_1$I$_2$ [Fig.~\ref{fig10}(d)] pairs. In fact, 
electrons scattered by the collective motion between I atoms will interfere destructively, instead 
of constructively, when $|h-k| \neq 3n$. 
This different response is attributed to the fact that CdI$_2$ (space group P$\bar{3}$m1) lacks
mirror symmetry with respect to the plane containing Cd atoms. %; this aspect is subject to further confirmation.  

\section{Conclusions} \label{sec.Conclusions}

In this manuscript we have benchmarked a new first-principles theory for the calculation of diffuse 
scattering in solids, as introduced first in the parallel paper, Ref.~[\onlinecite{arXiv:2103.10108}]. In a nutshell, we have demonstrated 
that our method can calculate efficiently and accurately multi-phonon scattering processes using as test cases 
bulk MoS$_2$ and bP, as well as several 2D materials. 

Starting from 2D MoS$_2$ we have validated our methodology by comparing successfully our results obtained 
within the LBJ and SDM theories. These theories enable one to calculate scattering patterns 
in a different fashion and at the same time justify the accuracy of each other. For completeness, we have 
explored in detail the formal mathematical link between the two theories. We emphasize that SDM is a broad 
approach with several applications in DFT and beyond~\cite{Zacharias_2020} 
which can be extended to study non-equilibrium dynamics~\cite{Seiler2021}; here we have simply demonstrated 
the physical significance of SDM in reproducing all-phonon diffuse patterns. We 
have also shown that the Einstein model fails completely in describing diffuse scattering, but it can
provide a good estimate for the contribution of multi-phonon interactions.

We further demonstrate our implementation of the all-phonon LBJ theory by evaluating 
scattering patterns of 2D transition-metal dichalcogenides (MoSe$_2$, WSe$_2$, and  WS$_2$), 
graphene, and 2D CdI$_2$. Remarkably, for 2D CdI$_2$ we find that multi-phonon processes 
contribute above 60\% to diffuse scattering. We clarify that this result should not be 
viewed as a limiting case, but rather a plausible outcome expected for several 
2D materials sharing similar Einstein phonon frequencies~\cite{Mounet2018}.

The present work helps to understand the quality of experimental measurements and investigate primary, or secondary, 
features in the scattering patterns of solids. For example, our results for bulk MoS$_2$ reveal that the measured
 diffuse background signals can be mainly explained by multi-phonon interactions. Furthermore, our multi-phonon 
calculations for bP demonstrate clearly the emergence of new primary features. % due to excitations of more than one phonons.  
Importantly, our finding suggests that extracting band-resolved phonon populations 
from the experimental data of bP by relying only on the one-phonon theory would be inaccurate.

Beyond studying the various phonon contributions to the diffuse patterns, we 
examine the scattering signatures arising from individual atomic and interatomic vibrational motions. 
Our analysis reveals that the collective displacement between specific pairs of atoms are responsible 
for the main fine structures observed experimentally. Clarifying the origin of these 
distinct features may help interpreting the data from a bonds perspective~\cite{Nicholson2018}, 
especially in materials with multiple atom species and/or multiple atoms per unit cell. 

We emphasize that our methodology creates a new framework in the interpretation of time-resolved electron, or X-ray,
experiments allowing for a reverse-engineering analysis to uncover transient phonon populations. In particular, one could 
combine the all-phonon scattering intensity with experimental data to single out multi-phonon contributions, and then 
extract phonon population dynamics using the strategy described in Ref.~[\onlinecite{Cotret_2019}]. 
This approach requires experimental data across multiple Brillouin zones extending to regions in reciprocal space where 
multi-phonon excitations can be dominant. We clarify that even if the occupancy of a 
single phonon mode is affected by photo-excitation in pump-probe experiments, the multi-phonon theory is 
still necessary to describe accurately the changes induced in diffuse scattering signals.

The present approach is as simple as efficient and can be implemented straightforwardly 
in any software package dealing with phonon properties of materials.
Given the generality of our methodology it should be possible to apply it in a large-scale
high-throughput manner for studying all-phonon diffuse scattering in solids. For systems experiencing 
a high-degree of anharmonicity, one could upgrade the phonons using 
the self-consistent harmonic approximation~\cite{Errea_2014,Patrick_2015}, or combine Eq.~\eqref{eq.intensity_T} with
{\it ab initio} molecular dynamics~\cite{Zacharias_2020_b}. For special cases, 
including (i) doped graphene~\cite{Lazzeri_2006}, or heavily boron doped diamond~\cite{Caruso_2017}, and
(ii) undoped semiconductors whose band gap energy is comparable to their phonon energies,
a breakdown of the adiabatic approximation is to be expected. In these cases, approaches beyond 
static density-functional perturbation theory and the frozen-phonon method are necessary to 
account for nonadiabatic phonon dispersions via the calculation of the phonon self-energy~\cite{Giustino_2017}.  

Electronic structure calculations performed in this study
are available on the NOMAD repository~\cite{nomad_doi}.

\acknowledgments

M.Z. acknowledges financial support from the Research Unit of Nanostructured Materials Systems (RUNMS)
and the program META$\Delta$I$\Delta$AKT$\Omega$P of the Cyprus University of Technology.
H.S. was supported by the Swiss National Science Foundation under Grant No.~P2SKP2\textunderscore184100.
F.C. acknowledges funding from the Deutsche Forschungsgemeinschaft (DFG) - Projektnummer 443988403. 
F.G. was supported by the Computational Materials Sciences Program funded by the U.S. Department of Energy, 
Office of Science, Basic Energy Sciences, under Award DE-SC0020129.
R.E. acknowledges funding from the European Research Council (ERC) under the European Union’s Horizon 2020
research and innovation program (Grant Agreement No. ERC-2015-CoG-682843) and by the Max Planck Society.
We acknowledge that the results of this research have been achieved
using the DECI resource Saniyer at UHeM based in Turkey~\cite{PRACE}
with support from the PRACE aisbl, computation time provided by the HPC Facility of the
Cyprus Institute~\cite{CyI}, and HPC resources from the Texas 
Advanced Computing Center (TACC) at The University of Texas at Austin~\cite{TACC}.

\begin{figure}[h!]
  \includegraphics[width=0.45\textwidth]{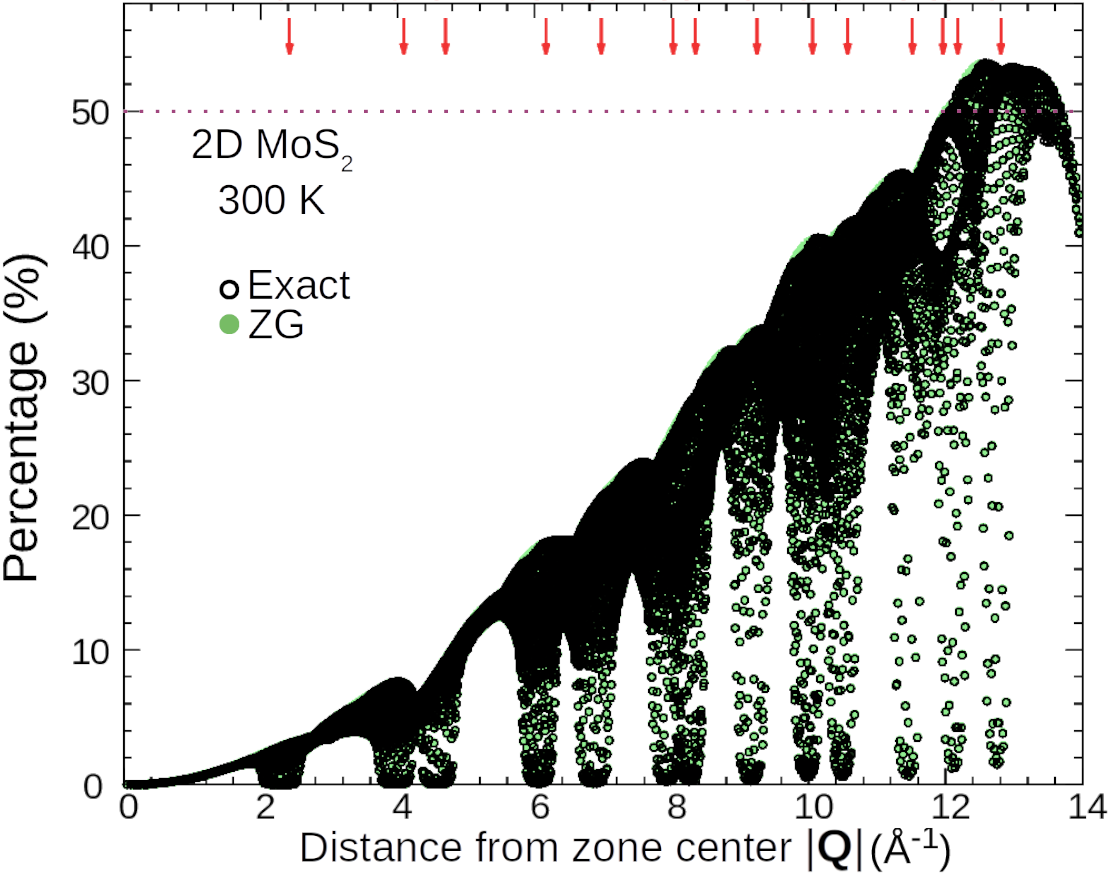}
  \caption{Percentage contribution of multi-phonon processes to the all-phonon 
  scattering intensity, $I_{\rm multi}/I_{\rm all}$, of monolayer MoS$_2$ 
  as a function of the scattering vector's distance $|\bQ|$ from the zone-center. 
  Black circles represent calculations using the exact formula in Eq.~\eqref{eqa1.7} 
  and a Brillouin zone {\bf q}-grid of size $50\times50$. The green discs represent 
  calculations using ZG displacements and a Brillouin zone {\bf q}-grid of size $300\times300$.
  Red arrows indicate distances for which $|\bQ|=|\bG|$, i.e. when Bragg scattering occurs. 
  The horizontal dashed line intersects the vertical axis at 50\%.
  \label{figA1} }
\end{figure}

\begin{appendix}
\numberwithin{figure}{section}
\begin{widetext}
\section{Equivalence between Eq.~\eqref{eq.ZG_struct_factor} and Eq.~\eqref{eqa1.7}} \label{app.equiv}

In this Appendix we show the equivalence between Eq.~\eqref{eq.ZG_struct_factor} and Eq.~\eqref{eqa1.7}. 
For the sake of clarity, we exclude from the discussion the terms arising from the phonons in group $\mathcal{A}$. 
This does not constitute a limitation, since the contribution of these terms vanishes in the thermodynamic limit~\cite{Zacharias_2020}.

We start the derivation with the aid of Eq.~\eqref{eq_key} and observe that the ZG scattering intensity can be written as: 
\begin{eqnarray}\label{eq.ZG_sf0}
 I_{\rm ZG}(\bQ,T) %&=& I^{\{\tau^{\rm ZG}\}}(\bQ) \nonumber \\
= \sum_{pp'} \sum_{\k \k'} f_\k (\bQ) f^*_{\k'} (\bQ) 
            e^{i \bQ \cdot [\bR_p - \bR_{p'} + \bt_\k - \bt_{\k'} ]} e^{-\frac{1}{2}  
  \big\{ \bQ \cdot  \big( \bDt^{\rm ZG}_{p\k } -  \bDt^{\rm ZG}_{p'\k' } \big) \big\}^2 },
\end{eqnarray}
Substituting Eq.~\eqref{eq.realdtau_method00} inside Eq.~\eqref{eq.ZG_sf0}
and performing some straightforward algebra yields: 
\begin{eqnarray}\label{eq.ZG_sf_2}
I_{\rm  ZG }(\bQ,T) =  \sum_{pp'} \sum_{\k \k'} f_\k (\bQ) f^*_{\k'} (\bQ)
    e^{ i \bQ \cdot [ \bR_p - \bR_{p'} + \bt_\k - \bt_{\k'} ] } 
     e^{-W_{\k} (\bQ,T)} \, e^{-W_{\k'} (\bQ,T)} \, 
     e^{ P_{pp',\k\k'} (\bQ,T)}e^{\Delta_{pp',\k\k'}(\bQ,T)},
\end{eqnarray}
where
\begin{eqnarray}\label{eqa1.8_b_pp}
 P_{pp',\k\k'} (\bQ,T)  &=& \frac{2 M_0 N^{-1}_p}{\sqrt{M_\k M_{\k'}}} \sum_{\bq \in \mathcal{B},  \nu }   u^2_{\bq \nu} 
\sum_{\a \a'}  Q_\a Q_{\a'}   
\text{Re}\bigg[ e_{\k\a,\nu} (\bq) e^{*}_{\k'\a',\nu} (\bq) e^{i\bq \cdot ({\bf R}_p - {\bf R}_{p'} )} \bigg]
\end{eqnarray}
and
\begin{eqnarray}\label{eq.error}
 \Delta_{pp',\k\k'}(\bQ,T) =
\frac{2 M_0}{N_p}
 \sum_{\a \a'}   Q_\a Q_{\a'} 
 \sum_{\substack {\bq \neq \bq' \in B \\ \nu \neq \nu'}}  
 &\Bigg[& - \frac{{\rm Re}\big[ e_{\k\a,\nu} (\bq) e^{i\bq \cdot {\bf R}_{p}}\big] {\rm Re} 
   \big[ e_{\k\a',\nu'} (\bq') e^{i\bq' \cdot {\bf R}_{p}} \big] }{M_\k}  - \k p \leftrightarrow \k' p'  \\ 
 &+& \frac{{\rm Re}\big[ e_{\k\a,\nu} (\bq) e^{i\bq \cdot {\bf R}_{p}}\big] {\rm Re} 
 \big[ e^*_{\k'\a',\nu'} (\bq') e^{i\bq' \cdot {\bf R}_{p'}} \big] }{\sqrt{M_\k M_{\k'}}}  
+ \k p \leftrightarrow \k' p' \Bigg] u_{\bq \nu} u_{\bq' \nu'}  S_{\bq \nu} S_{\bq' \nu'}. \nonumber
\end{eqnarray}
The function $\Delta_{pp',\k\k'}(\bQ,T)$ represents the deviation from the exponents of the Debye-Waller and 
phononic factors. 
The notation $ \k p \leftrightarrow \k' p'$ indicates the previous term with the indices $\k$, $p$ and $\k'$, $p'$ interchanged. 
We now take the Taylor expansion of $e^{\Delta_{pp',\k\k'}(\bQ,T)}$
and, for simplicity, we keep only terms up to second order in atomic displacements to obtain:  
\begin{eqnarray}\label{ZG_sf_3}
I_{\rm ZG}(\bQ,T) &=&  \sum_{pp'}  \sum_{\k \k'} f_\k (\bQ) f^*_{\k'} (\bQ) 
            e^{i \bQ \cdot [\bR_p - \bR_{p'} +  \bt_\k - \bt_{\k'} ]} 
            e^{-W_{\k} (\bQ,T)} \, e^{-W_{\k'} (\bQ,T)} \,
            e^{ P_{pp',\k\k'} (\bQ,T)} \nonumber \\ %I_{\rm all}(\bQ,T)  
            &+&  \sum_{pp'}  \sum_{\k \k'} f_\k (\bQ) f^*_{\k'} (\bQ) 
            e^{i \bQ \cdot [\bR_p - \bR_{p'} +  \bt_\k - \bt_{\k'} ]} 
            e^{-W_{\k} (\bQ,T)} \, e^{-W_{\k'} (\bQ,T)} \,
             \Delta_{pp',\k\k'}(\bQ,T). 
\end{eqnarray}
In view of translational symmetry of the lattice, the first line of the
 above relation gives exactly the all-phonon term, $I_{\rm all}(\bQ,T)$, as given by Eq.~\eqref{eqa1.7}.
The second line is recognized as the leading error in the evaluation of the ZG scattering intensity.
Now we substitute Eq.~\eqref{eq.error} into Eq.~\eqref{ZG_sf_3}, perform the summations over $p$ and $p'$ using twice 
the relation $\sum_{p} e^{i ( \bQ - \bq )\cdot {\bf R}_p} = N_p \, \delta_{\bQ,\bq +\bG}$,
and apply time-reversal symmetry, i.e. $I_{\rm ZG}(\bQ,T) = I_{\rm ZG}(-\bQ,T) $. Hence, 
the ZG scattering intensity simplifies to: 
\begin{eqnarray}\label{ZG_sf_4}
I_{\rm ZG}(\bQ,T) &=& I_{\rm all}(\bQ,T)  
   + \sum_{\k \k'} f_\k (\bQ) f^*_{\k'} (\bQ) \cos\big[\bQ \cdot (\bt_\k - \bt_{\k'}) \big]
e^{-W_{\k} (\bQ,T)} \, e^{-W_{\k'} (\bQ,T)} \, \Delta_{\k\k'}(\bQ,T), 
\end{eqnarray}
where the error term $\Delta_{\k\k'}(\bQ,T)$ is given by: 
\begin{eqnarray}\label{ZG_sf_5}
\Delta_{\k\k'}(\bQ,T) =
   &-& 2 \frac{ M_0 N_p}{M_\k} \sum_{ \a \a' } \, Q_\a Q_{\a'}  \delta_{Q,G}
    \sum_{\substack {\bq \in \mathcal{B} \\ \nu < \nu'}} 
    \text{Re} [e_{\k\a,\nu}(\bq) e^{ *}_{\k\a',\nu'}(\bq)]
     u_{\bq \nu} u_{\bq \nu'}  S_{\bq \nu} S_{\bq \nu'} \nonumber \\  
   &-& 2\frac{M_0 N_p}{M_{\k'}} \sum_{ \a \a' } \, Q_\a Q_{\a'}  \delta_{Q,G}
   \sum_{\substack {\bq \in \mathcal{B} \\ \nu < \nu'}} 
    \text{Re} [e_{\k'\a,\nu}(\bq) e^{ *}_{\k'\a',\nu'}(\bq)]
     u_{\bq \nu} u_{\bq \nu'}  S_{\bq \nu} S_{\bq \nu'} \nonumber \\ 
   &+& 4 \frac{M_0 N_p}{\sqrt{M_\k M_{\k'}}} \sum_{ \a \a' } \, Q_\a Q_{\a'}  
   \sum_{\substack {\bq \in \mathcal{B} \\ \nu < \nu'}} 
    \text{Re} [e_{\k\a,\nu}(\bq) e^{*}_{\k'\a',\nu'}(\bq)]
    u_{\bq \nu} u_{\bq \nu'}  S_{\bq \nu} S_{\bq \nu'}.  
\end{eqnarray}
The first and second lines of the above expression are associated
with the error in the evaluation of diffuse scattering for $\bQ=\bG$. 
By comparing now Eq.~\eqref{eq.minimiz_function} with Eq.~\eqref{ZG_sf_5}, it is evident that 
$\Delta_{\k\k'}(\bQ,T)$ is minimized together with $E(\{S_{\bq  \nu}\},T)$ owing to the 
choice of signs made for the ZG displacement.
The same arguments also apply for the elimination of the error arising beyond  
second order in atomic displacements, i.e. terms including higher powers of $\Delta_{pp',\k\k'}(\bQ,T)$.
This completes the proof that Eq.~\eqref{eq.ZG_struct_factor} and Eq.~\eqref{eqa1.7} are equivalent in 
the thermodynamic limit. As a numerical demonstration, in Fig.~\ref{figA1} we show that multi-phonon 
contributions calculated with Eq.~\eqref{eq.ZG_struct_factor} and Eq.~\eqref{eqa1.7} are, indeed, identical. 

\end{widetext}

\end{appendix}

\bibliography{references}

\end{document}